\documentclass[apj]{emulateapj}
\journalinfo{\textsc{Accepted in The Astrophysical Journal}}
\slugcomment{Received 2015 May 29; Accepted 2015 October 6}
\usepackage{url}

\usepackage{apjfonts}
\usepackage{natbib}
\usepackage{booktabs}
\usepackage{threeparttable}

\usepackage[colorlinks,
        hypertexnames=false,
        pagebackref=true,
        linkcolor=blue,    
    citecolor=blue,        
    filecolor=magenta,     
    urlcolor=blue          
]{hyperref}                
\usepackage{amsmath}

\newcommand{\nar}{{\it New Astr. Rev.}}

\shorttitle{Sun-Earth Connection of an Earth-Directed CME Magnetic Flux Rope}
\shortauthors{Vemareddy and Mishra}

\begin{document}
\title{A Full Study on the Sun-Earth Connection of an Earth-Directed CME Magnetic Flux Rope}
\author{P.~Vemareddy$^1$ and W.~Mishra$^{2}$}
\affil{$^1$Indian Institute of Astrophysics, II Block, Koramangala, Bangalore-560 034, India}
\affil{$^2$Department of Geophysics and Planetary Sciences, University of Science and Technology of China, Hefei-230026, China}
\email{vemareddy@iiap.res.in, wageesh@ustc.edu.cn}
\begin{abstract}
We present an investigation of an eruption event of coronal mass ejection (CME) magnetic flux rope (MFR) from source active region (AR) NOAA 11719 on 11 April 2013 utilizing observations from SDO, STEREO, SOHO, and WIND spacecraft. The source AR consists of pre-existing sigmoidal structure stacked over a filament channel which is regarded as MFR system. EUV observations of low corona suggest a further development of this MFR system by added axial flux through tether-cutting reconnection of loops at the middle of sigmoid under the influence of continuous slow flux motions during past two days. Our study implies that the MFR system in the AR is initiated to upward motion by kink-instability and further driven by torus-instability. The CME morphology, captured in simultaneous three-point coronagraph observations, is fitted with Graduated Cylindrical Shell (GCS) model and discerns an MFR topology with orientation aligning with magnetic neutral line in the source AR. This MFR expands self-similarly and is found to have source AR twist signatures in the associated near Earth magnetic cloud (MC). We further derived kinematics of this CME propagation by employing a plethora of stereoscopic as well as single spacecraft reconstruction techniques. While stereoscopic methods perform relatively poorly compared to other methods, fitting methods worked best in estimating the arrival time of the CME compared to in-situ measurements. Supplied with values of constrained solar wind velocity, drag parameter and 3D kinematics from GCS fit, we construct CME kinematics from the drag based model consistent with in-situ MC arrival.
\end{abstract}

\keywords{Sun:  heliosphere--- Sun: flares --- Sun: coronal mass ejection --- Sun: magnetic fields---
Sun: filament --- Sun: solar-terrestrial relations}
\section{Introduction}
\label{Intro}
Coronal Mass Ejections (CMEs) are magnetically driven gigantic events whose disturbance in the outer corona influences the space-weather to a wide range. When propagating from the Sun, they appear to have three-part structure in white light \citep{illing1985,burlaga1988, forsyth2006} with leading edge, core and cavity. Although not quite clear in the observations, the cavity is supposed to have its connections in the source regions on the Sun and is approximated by a coherent, large scale magnetic magnetic flux rope (MFR), manifested by twisted magnetic field lines. A recent statistical study suggests that at least 40\% of CMEs observed by space borne instruments have a clear MFR structure \citep{vourlidas2013}. Many studies from space and ground based observations suggested that this MFR structure in the outer corona is an evolved form of filaments seen in H$\alpha$ or sigmodal structure in soft X-rays in the so called magnetic active regions (e.g., \citealt{lepping1990, mckenzie2008, vemareddy2011,zhangj2012,howard2012,chengx2012}) on the Sun. Central to the space-weather phenomena and the Sun-Earth connections of the CMEs having signatures of this predicted MFR topology, a majority of past and current scientific research dwells on basic questions like how such MFR like structures formed or comes into existence (e.g., \citealt{pneuman1983, ballegooijen1989, fany2001, fany2004, gibson2004, gibson2006, leka1996}), what are the initiating conditions in the source active region (e.g., \citealt{moore2001, priest2002, linj2003, forbes2006, fany2007, kliem2006, martin1985, antiochos1999, chenpf2011, chengx2012, vemareddy2011, vemareddy2014}) and what kind of MFR evolution drive its outward propagation (e.g., \citealt{temmer2012, temmer2011, gopalswamy2000, gopalswamy2001, vrsnak2002, manoharan2006, vrsnak2013, davies2009, webb2009, forsyth2006, harrison2012}) in the extended corona and interplanetary medium.

Two strong ideas that the past and current observations on the Sun implied were emergence of the MFR from beneath the inner photosphere and formation of the MFR from sheared arcade. In the process of active region formation while it emerges, the MFR structure would emerge as a bipolar region and equilibrates with the overlying, pre-existing structure \citep{fany2009}. Numerical simulations of the bodily emergence of the MFR \citep{fany2001, fany2003, magara2001, gibson2004,okamoto2009, lites2010} showed the formation of sheared arcade in the corona after the MFR axis reached the photosphere. In the case of fully emerged active region, the equilibrated MFR structure in the form of sheared arcade evolves to full fledged MFR. Based on these ideas, different theoretical and observational models have been proposed and constructed \citep{antiochos1998, antiochos1999, linj2003, forbes2006, kliem2006, fany2007,chenpf2011}. In all of them, twisted MFR or sheared arcade which is a source of magnetic helicity, is the basic ingredient of pre-eruptive magnetic configuration. \citet{pneuman1983} showed how helical MFRs can be formed by reconnection in a sheared coronal arcade. During the progressive reconnection phase, the magnetic flux gets canceled, transforming the magnetic field lines in the sheared arcade along the magnetic inversion line to winding, helical field lines about a common axis of nearly formed MFR \citep{green2009, green2011}. Later, \citet{ballegooijen1989} proposed a mechanism for the formation of helical field lines from the reconnection of less sheared arcade along the polarity inversion line and also to build up the MFR. Submerging motion of fluxes about the PIL also plays vital role in bringing less sheared to highly sheared arcade. Usually the reconnection location is associated with the photospheric flux cancellation \citep{yurchyshyn2001, bellotrubio2005} and is evidently reported to be involved with the MFR formation and its eruption \citep{green2011}. 

In the low corona, triggering and driving of MFR eruption is another important step in the eruption process. Filament, prominence or sigmoid eruptions modeled by the numerical and theoretical constructions of MFRs have been successful in many respects of observed eruption features. The physics of these  models essentially based on loss of equilibrium of MFR after reaching a critical height \citep{forbes1991, priest2002}, kink instability due to exceeding MFR's twist \citep{torok2004,torok2005},  torus instability \citep{kliem2006, aulanier2010} of the MFR when there is a rapid decline of the background field in the direction of the expansion of the MFR. 

In the case of source active regions located near the disc center, the CME MFR eruption is very likely directed toward the Earth causing direct space weather impacts. In order to estimate the arrival times of CMEs to Earth, a vital step is to understand the kinematic evolution of CMEs while propagating through the heliosphere. Based on \textit{Large Angle and Spectrometric Coronagraph} (LASCO; \citealt{brueckner1995}) coronagraph observations on board the \textit{Solar and Heliospheric Observatory} (SOHO), many studies found that CME having fast speed near the Sun, decelerates during its journey to the Earth while CME having slow speed accelerates \citep{lindsay1999, gopalswamy2000, gopalswamy2001, yashiro2004, manoharan2006}. This finding clearly emphasizes the interaction of CME with the ambient solar wind medium. LASCO observes only out to $30R_{\odot}$ and can provide only projected kinematics of the CME while CME speed changes significantly beyond the FOV of coronagraphs. Therefore two point measurements of speed, one near the Sun and another near the Earth, are not sufficient to fully capture the physics of CME evolution in the heliosphere. 

In the present era, the estimation of CME kinematics from its lift-off in the low corona to the Earth or even beyond is possible with multiple views (A and B) of the Sun-Earth space from the \textit{Solar Terrestrial RElations Observatory} (STEREO; \citealt{kaiser2008}). Using coronagraphs (CORs) and heliospheric imagers (HIs) of \textit{Sun-Earth Connection Coronal and Heliospheric Investigation} (SECCHI; \citealt{howard2008}) onboard STEREO, several studies have been carried out attempting to associate remote imaging observations with in situ observations near the Earth \citep{davies2009,mostl2010, liuying2010, harrison2012, temmer2012, lugaz2012, mishra2013,liuying2014b,rollett2014,mostl2015}. Single and/or multiple spacecraft reconstruction methods on HI observations have shown major limitation in estimating kinematics and the inherent difficulties involved in interpreting the HI observations of the CMEs.  However, most of these studies have been carried out mainly for Earth-directed CMEs during STEREO separation was small. Therefore, the tracking of the CME or its associated MFRs moving away form the observer seems to be rarely undertaken for study \citep{liuying2014a}. Hence, the assessment of relative performance of reconstruction methods with different geometry, assumed for a CME receding from the observer, is an obvious next step for the solar-terrestrial scientist.

In the present study, we focused on an eruption event of April 11, 2013 from the NOAA AR 11719, uniquely involving its gradual build up and eventual eruption to Earth directed CME. The initiation mechanism based on morphological twist signatures of this MFR event was extensively studied in \citet[hereafter VZ14]{vemareddy2014}. Interestingly, the solar energetic particle (SEP) event on 11 April, 2013 was found to originate from this eruption event \citep{lario2014} which produced first large Fe-rich SEP event of solar cycle 24 \citep{cohen2014}.  In view of further Sun-Earth connection of this CME MFR event, in this manuscript, we extended the study of VZ14 on the formation or development scenario of MFR in the source AR and its propagation in the inner heliosphere after its eventual eruption from the source AR. Utilizing detailed multi-wavelength EUV observations from \textit{Atmospheric Imaging Assembly} (AIA; \citealt{lemen2012}), we study the evolution history of the source AR prior to three days. In concurrent with the observed morphological evolution, we analyzed vector magnetograms obtained from \textit{Helioseismic and Magnetic Imager} (HMI; \citealt{schou2012}) to support or revert the scenario of MFR formation by canceling magnetic flux in the source active region. Moreover, taking advantage of multi-view observations from STEREO/SCCHI, SOHO/LASCO coronagraphs, we determined orientation of the underlying MFR by a parametrized fitting model to the observed white light CME morphology. We further tracked this CME, which is moving away from the STEREO observer, and estimated its kinematics using stereoscopic and single spacecraft reconstruction methods. The estimated arrival time of CME MFR at L1 is discussed in comparison with in situ plasma parameters to assess the reliability of the employed tracking methods. CME Kinematics are also derived and discussed using drag based model. Further,  the AR twist signatures are explored from in situ magnetic field observations of this CME associated magnetic cloud. Such a study involving multi-scale observations examines the complete Sun to Earth connection of an event and furthers the understanding in space-weather phenomena.

The manuscript is structured as follows: in section~\ref{sec2}, the requirement of observations from various instruments is outlined, and in section~\ref{sec3}-\ref{sec4}, the formidable conditions for the formation/development are explored using coronal EUV and photospheric magnetic field observations. For completeness, the details of onset and driving mechanisms of this MFR eruption (VZ14) are given in section~\ref{sec5}. The orientation of the MFR while expanding from the source active region is determined in section~\ref{sec6}. Its propagation kinematics toward Earth and interplanetary twist signatures are presented in section~\ref{sec7}. We concluded with a summarized discussion of this investigation in section~\ref{sec8}. 
\begin{figure*}
\centering
\includegraphics[width=.85\textwidth]{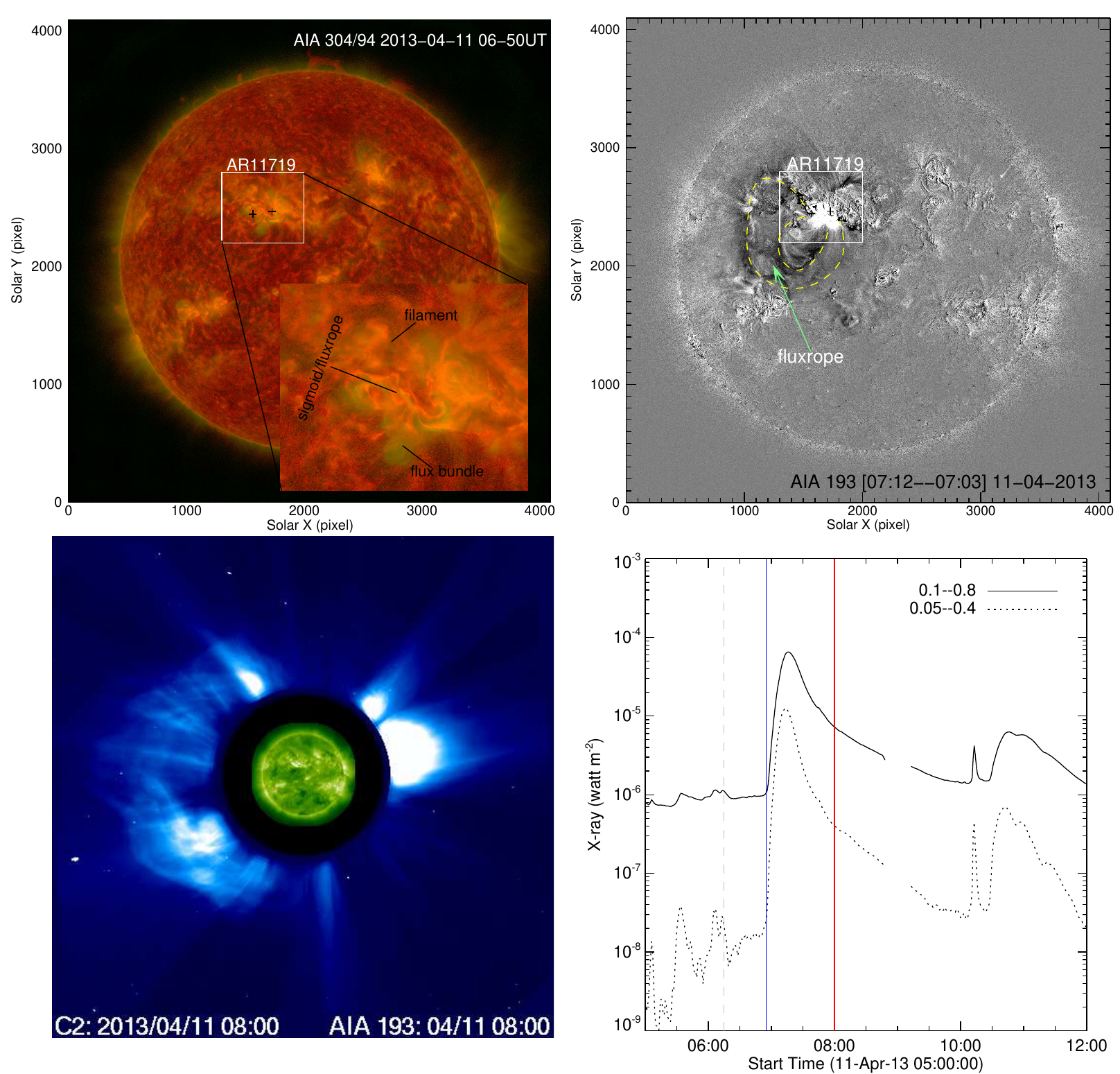}
\caption{\textbf{Top left:} Composite image of the solar disk prepared from AIA 304 and 94A observations. The source region of CME on the Sun is marked with rectangular inset. The pre-eruptive AR 11719 consisted of sigmoidal structure that deviates from the filament channel. \textbf{Top right:} Difference image (07:12-07:03UT) of AIA 193\AA~indicates EUV disturbance and outlines the rapidly expanding MFR from AR11719. \textbf{Bottom Left:} Propagating CME and bright leading edge in C2 field of view on board SoHO. Image of the Sun from AIA 193\AA~observation is also placed at the occulter disk position.  \textbf{Bottom Right:} time profile of GOES x-ray flux in low and high frequency bands. The peak flux corresponds to M6.5 flare that commenced at 06:55UT (vertical green line) subsequent to the onset of the MFR at 06:15UT (dashed grey vertical line). The red vertical line indicates the time of the captured C2 image shown in the left panel. }
\label{Fig1}
\end{figure*}

\section{Outline of Observations}
\label{sec2}
The overall eruption event of our interest is outlined in Figure~\ref{Fig1}. The observations of the source active region NOAA 11719, both at photospheric and low corona heights, are well covered by the Solar Dynamic Observatory and its component instruments AIA and HMI. Using AIA channels, especially 304 and 94\AA~observations (0.6 arcsec/pixel), the pre-eruptive AR [at 06:50UT] magnetic flux system is noticed to present an inverse-S sigmoidal structure (top left panel) that is stacked over filament channels. During the initial upward rise motion at 06:15UT (just before the main eruption around 06:55UT), the plasma emission (in 131 and 94\AA~observations) along this sigmoidal structure supports the interpretation that the sigmoid is manifested by a coherent core twisted flux system surrounded by twisted overlying flux bundle (VZ14). We regard this coherent core flux along the sigmoid as main body of the MFR, which is wound by the overlying flux bundle. The time difference image of 193\AA~(top right panel) shows an EUV disturbance in the low corona during this main phase of eruption. This EUV disturbance also outlines the expanding MFR having legs rooted from the pre-determined ("+") sigmoid legs. The eruption of the sigmoid MFR from the low corona is preceded by a GOES M6.5 flare (bottom right panel). Importantly, its CME-driven shock propagates at high altitudes above the solar surface and was likely the source of the solar energetic particles observed near Earth \citep{lario2014, cohen2014}. 

This propagating CME in the outer corona was well observed by the \textit{Large Angle and Spectrometric Coronagraph} (LASCO; \citealt{brueckner1995}) on board the \textit{Solar and Heliospheric Observatory} (SoHO) and the SECCHI on board the STEREO (A and B). The CME structure emerged out of the occulter to become a halo like structure at around 08:30UT. Figure~\ref{Fig1}(bottom left) of LASCO/C2 image discerns a bright annular structure on East limb of the disc resembling the leading edge followed by invisible cavity of the MFR. While expanding, the transit of the MFR through multiple heights, in the field-of-views (FOVs) of COR1 (1.1-4$R_{\odot}$), COR2 (2-15$R_{\odot}$), LASCO/C2 (1.5-6$R_{\odot}$), LASCO/C3(3.5-30$R_{\odot}$) situated at perspective viewing angles, uniquely show a direct connection between the heliospheric MFRs and the sigmoidal structures in the source active region (more details in Section~\ref{sec6}). The majority of observed CMEs in the past solar cycle had clear MFR structures and no presently known physical mechanism can produce a large-scale fast eruption beyond 10$R_{\odot}$ without ejecting a MFR \citep{vourlidas2013}. Having these supporting observational signatures of MFR topology both in the source region sigmoid and CME morphology, we use the term MFR while describing and interpreting the observations in its favor. The CME propagation in the heliosphere was tracked by HI-I (15-90$R_{\odot}$) and HI-II (70-330$R_{\odot}$) instruments of SECCHI. In situ plasma and magnetic field observations are obtained from WIND space mission. Taking advantage of these detailed observations, we focus on the objectives mentioned in the last section.  

\begin{figure*}
\centering
\includegraphics[width=.97\textwidth]{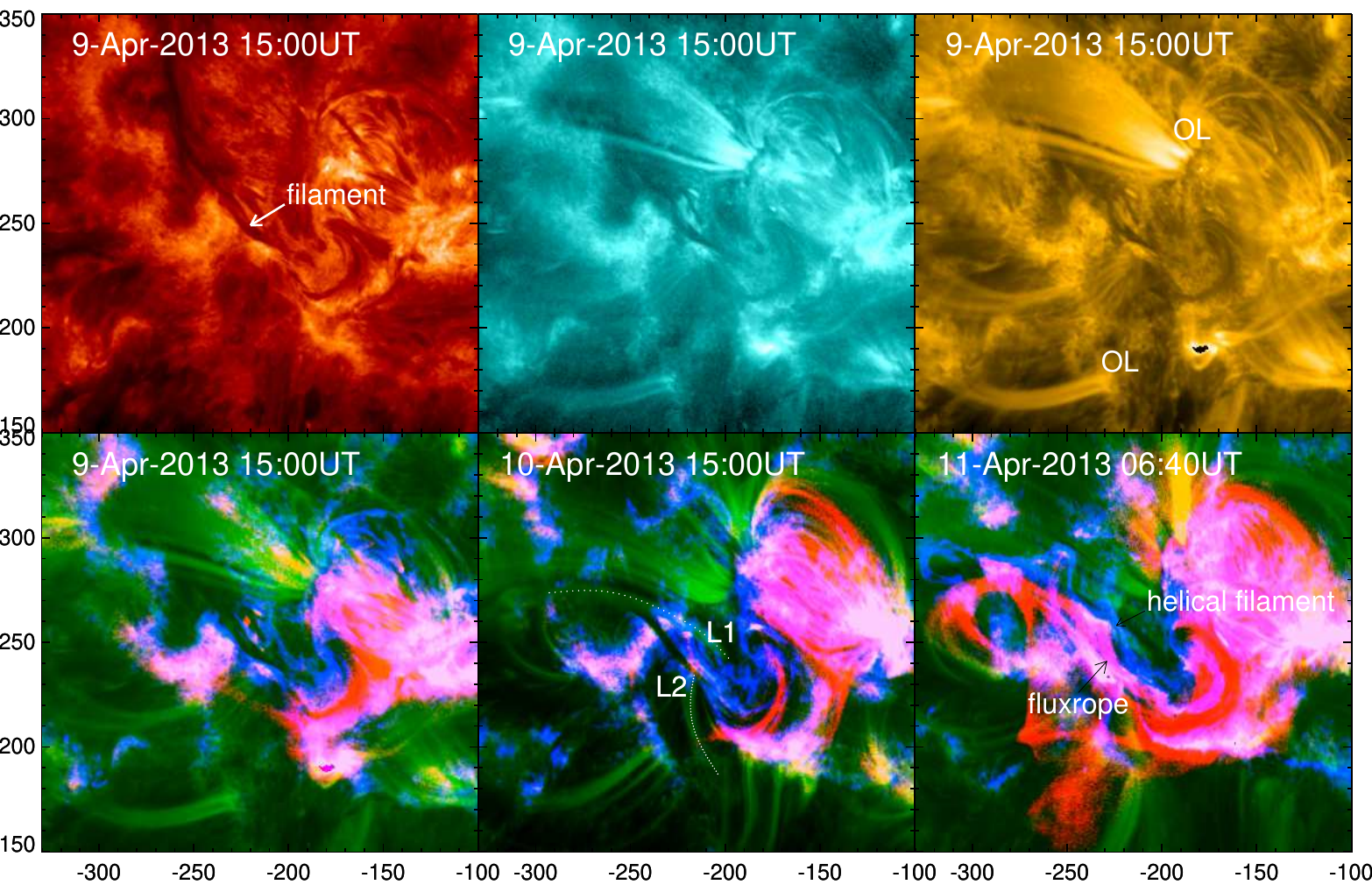}
\caption{The observational scenario of inverse-S sigmoidal MFR in AR 11719. \textbf{Top row:} The observations of the corona in AIA  304, 94, 171\AA~ passbands on 9 April 2013. Note the distant loops (OL) overly the cold filament, \textit{Bottom row:} composite images of 304, 171, and 94\AA~referring to presence of loops overlying the filament and the developing sigmoidal MFR. Low lying loops L1 and L2 about the sigmoid and their evolution over the long time infers a scenario of tether-cutting reconnection after which they align the MFR axis. All image panels are in heliographic coordinates in arcsec units.}
\label{Fig2}
\end{figure*}

\begin{figure}[h!]
\centering
\includegraphics[width=.49\textwidth]{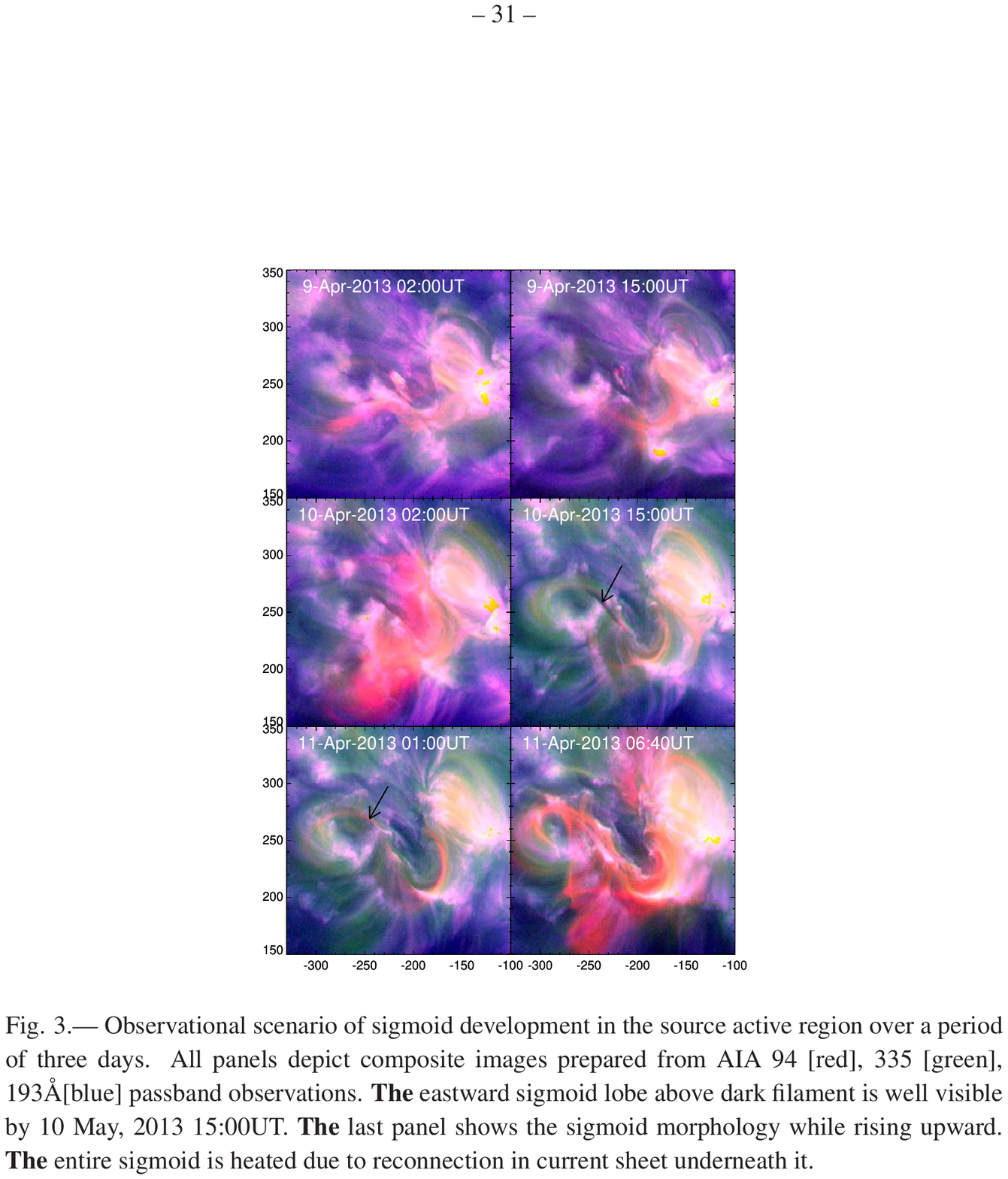}
\caption{Observational scenario of sigmoid development in the source active region over a period of three days. All panels depict composite images prepared from AIA 94 [red], 335 [green], 193\AA [blue]~passband observations. The eastward sigmoid lobe above dark filament is well visible by 10 May, 2013 15:00UT. The last panel shows the sigmoid morphology while rising upward. The entire sigmoid is heated due to reconnection in current sheet underneath it.}
\label{Fig3}
\end{figure}

\section{Development of MFR Structure: AIA Observations}
\label{sec3}
\begin{figure*}[!ht]
\centering
\includegraphics[width=.97\textwidth]{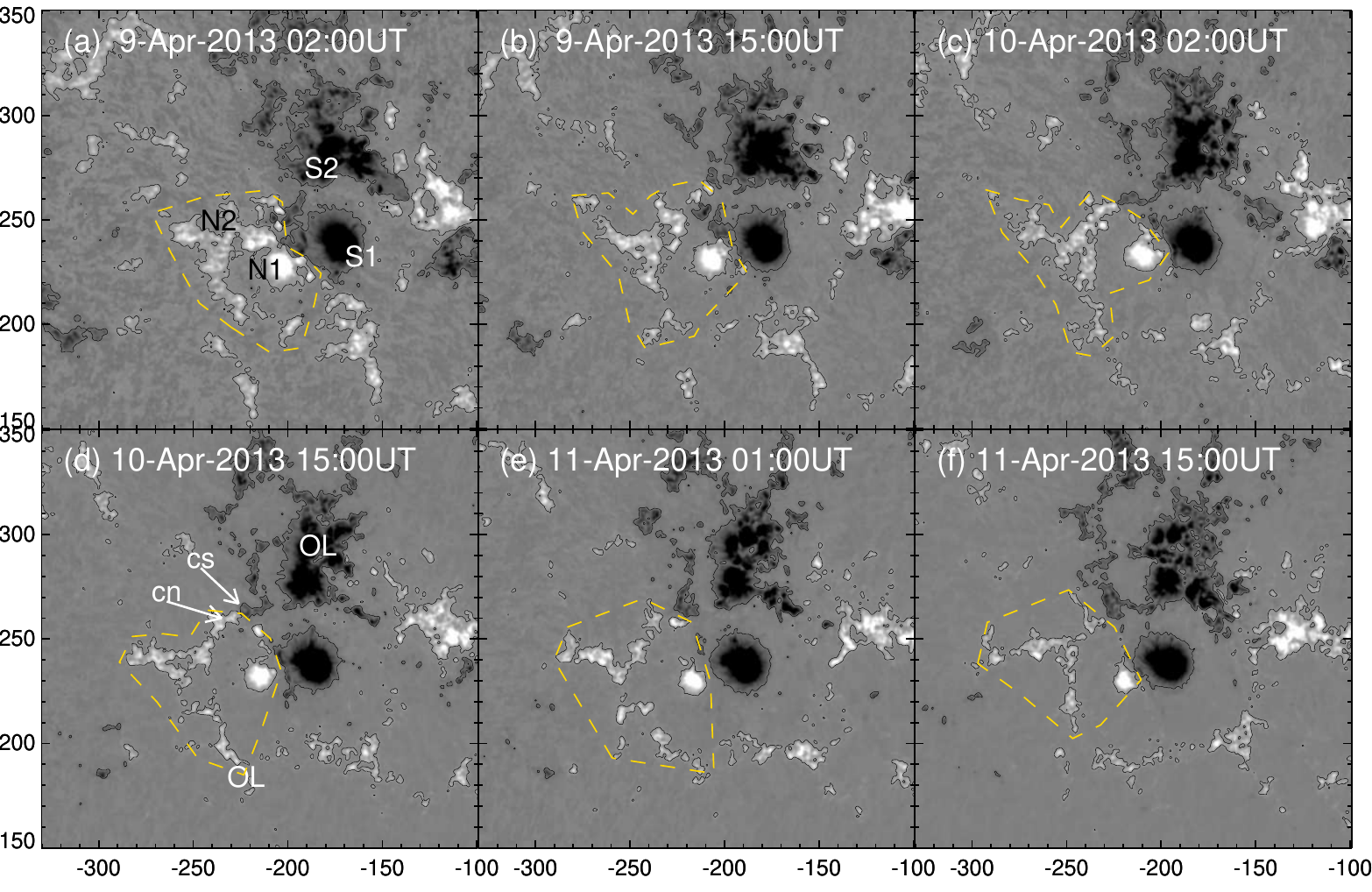}
\caption{Line-of-sight magnetic field observations from HMI corresponding to AIA coronal observations in Figure~\ref{Fig3}. Contours of $\pm120$G are overdrawn in each panel for highlighting magnetic regions from surrounding small scale values. Positive flux in the enclosed region of yellow (dashed) contour corresponds to East lobe leg of sigmoid and possibly involved in the course of development of sigmoidal MFR. Note that flux distribution smears away throughout three day evolution within this contour. Approaching and canceling flux features in panel (d) discern a possible scenario of reconnection of inclined loops inferred in Figure~\ref{Fig2}.}
\label{Fig4}
\end{figure*}

The active region 11719 contains a well observed filament from its appearance on the eastern limb of the solar disc. The filament consists of two sections which together give an inverse S-shape to it. These coronal features are identified to be associated with magnetic regions ({labels in Figure~\ref{Fig4}(a)}) at the photosphere. The lower section originates from main sunspot (S1) and has a cusp shape at the joining to upper section. Long coronal loops having footpoints from flux region S2 are well captured in AIA passbands, predominantly in 171\AA~and are observed to overly the filament, with conjugate footpoints lying in N1. In hotter passbands (2-10MK) of AIA especially in 94, 131\AA, the entire active region structure becomes complicated as further features of loops filled with hot plasma are observed. However, the surrounding structures from other parts of the flux regions  have similar thermal conditions. The field line system around the main filament is not quite enough to reveal the underlying topological configuration in detail. Because this entire configuration evolves under quasi-static conditions, we observe no significant large scale structural change for hours. 
 
The AIA filter passbands have a wide range of visibility of thermal plasma from chromosphere ($LogT=10^4K$) to hot, diffused corona ($\approx10$MK). In order to disclose details of connection between low lying filament and the overlying loops, we make use of composite images of different combinations prepared from AIA passbands. For a simultaneous view of corona from chromosphere, we showed in Figure~\ref{Fig1}(top left), the 304\AA~passband image at 06:50UT on April 11, 2013 blended with the image of 94\AA~passband at the same time. From this, it is obvious that the low lying filament channel in 304\AA~deviates in projection with the overlying hot sigmoid. The filament, the sigmoid above it and the overlying flux bundle demonstrate a complex flux systems stacked over each other. 

A similar observational view persists well before on 9 April in the 304, 131 and 171\AA~channels (Figure~\ref{Fig2}). The cool filament channel is overlaid by long plasma loops (OL) having their photospheric connections on either side of it. This is analyzed in successive composite frames prepared from 304, 171, and 94\AA~(as shown in 2nd row panels of Figure~\ref{Fig2}). The central part of the filament is increasingly surrounded by hot plasma   illuminated in 131\AA, which indicates the augmenting scenario of the main body of the MFR by wrapped field lines. This would be likely due to reconnection in the current sheets formed in the interface of MFR and surrounding field, by which the thermal energy heats the plasma to observable temperatures \citep{gibson2006}.
 
The observations are of three-dimensional structures projected onto the plane of the sky. Moreover, although plasma and field are frozen together in highly conducting corona so that coronal plasma structures essentially trace the magnetic field, plasma emission is the only possible distinct observable under certain density and temperature conditions dictated by an observing instrument. In this case, we made use of composite images prepared from different combinations of AIA channels (for quiet and flaring corona with combinations 94, 335, 193\AA; flaring corona, chromospheres, transition region in combinations of 131, 94, 304\AA) for our interpretations. They provide substantial evidence for the developing scenario of the sigmoidal structure during three days of evolution (Images not shown here).

The augmentation scenario of the MFR appears to occur in two stages (Figure~\ref{Fig2}). In the first stage, the long overlying loops, (OL) in the form of arcade, significantly turn to sheared arcade and then transform toward the east and west lobes of the sigmoid structure.  In the second phase, these increasingly inclined loops turn towards two inverse J-sections having footprints lying side by (L1 and L2) the filament. Evidently, the loops L1 and L2 reflect a probable involvement in traditional tether-cutting reconnection \citep{moore2001}. Once reconnection of these L1 and L2 loops occurs, formation of continuous sigmoid is inevitable, by added axial fluxtubes. These two phases possibly occur in a slow dynamical evolution, where infinitesimal displacements of the flux regions having these loops, induce continuous reconnection, which adds twisted flux threads wounding some common axis of the existing MFR. This mechanism therefore is identical to the sheared arcade to rope transformation proposed by \citet{ballegooijen1989}. Indeed, the observed magnetic flux distribution well supports this augmentation scenario which shows the canceling and approaching opposite flux patches side by the filament (Figure~\ref{Fig4}). Our interpretations are further supported by another set of observations (Figure~\ref{Fig3}) below.

Earlier on 9 April (Figure~\ref{Fig3}), the existence of the sigmoidal structure is not clear in hot channels except faintly observed filament channel. However, emission from the lower lobe section (like inverse J) is conspicuously present most of the time. After a day of evolution i.e., by April 10, a developing sigmodal structure becomes apparent with a well visible East lobe. Specifically, the hot overlying flux bundle at the middle of the sigmoid and its geometric evolution is of interest (VZ14). While undergoing dynamical evolution, this whole structure brightens intermittently, probably corresponding to small scale reconnections of the loop systems. One such major bright emission corresponds to a partial/failed eruption (around 18:00UT on April 10) in which the MFR lifted and appears to split so that the upper part expels and lower section settled back to its stable position \citep{gilbert2001, gilbert2007}. Data gaps around this period constrain to reveal more details on this partial eruption. After this, the two J-shaped sections merged to form a continuous sigmoidal MFR which is very apparent (panels at 15:00/10, 01:00/11 April), by this time the magnetic flux distribution shows fragmented and disappearing positive flux corresponding to the northern lobe structure of the MFR. 

Further, during the evolution of three days, it is difficult to interpret the magnetic topology of this flux system of filament/sigmoid from the projected imaging information. Bald patch seperatrix surface (BPSS) and hyperbolic flux tube (HFT) are two kinds of topological configurations associated with theoretically predicted MFR models \citep{titov2002}.  As suggested by van Ballagooijen, sigmoids formed from sheared arcades by photospheric flux cancellation driven by converging motions, which is the case here, are more likely associated with BPSS rather than HFT. When the sigmoid structure is regarded as MFR lying low in height, its middle section dips into dense photosphere and appears as two J-sections in the coronal observations like x-ray or EUV. However, the associated BPSS topology may turn to HFT while slow upward rise motion commences \citep{gibson2006} and continuous structure of MFR will be envisioned. Because fast reconnection occurs in a thinning current sheet with HFT, the MFR lifts off the filament completely with full eruption and is very unlikely for the eruption to be suppressed at that stage.

\begin{table}[h!]
\caption{Net Magnetic Flux from the contour region in Figure~\ref{Fig4}}
\begin{tabular}{l l l l}
\hline
Date [UT] 			&		axial flux  													 	&   total abs flux 													&  ratio [\%] \\
											&		[$10^{21}$ Mx] 									&   [$10^{21}$ Mx]												&  								\\
\hline	
April 9, 02:00   	&  	4.1  																& 	28.9																			&			14.1		\\
April 9, 15:00    	& 	3.1  																&	22.5																				&			13.7		\\
April 10, 02:00  	& 	2.9 															 &	22.7																			 &			13.0		\\
April 10, 15:00 	& 	2.3  															&	17.4																				&			13.3  	\\
April 11, 01:00  	& 	2.4  															&	17.3																				&			13.6		\\
April 11, 15:00  	& 	1.5  															&	16.5																				&			 9.1			\\
\hline
		\end{tabular}
	\label{tab1}
\end{table}

\section{Cancellation of Magnetic Flux and Build-up of Twist: HMI Observations}
\label{sec4}
\begin{figure}[h!]
\centering
\includegraphics[width=.48\textwidth]{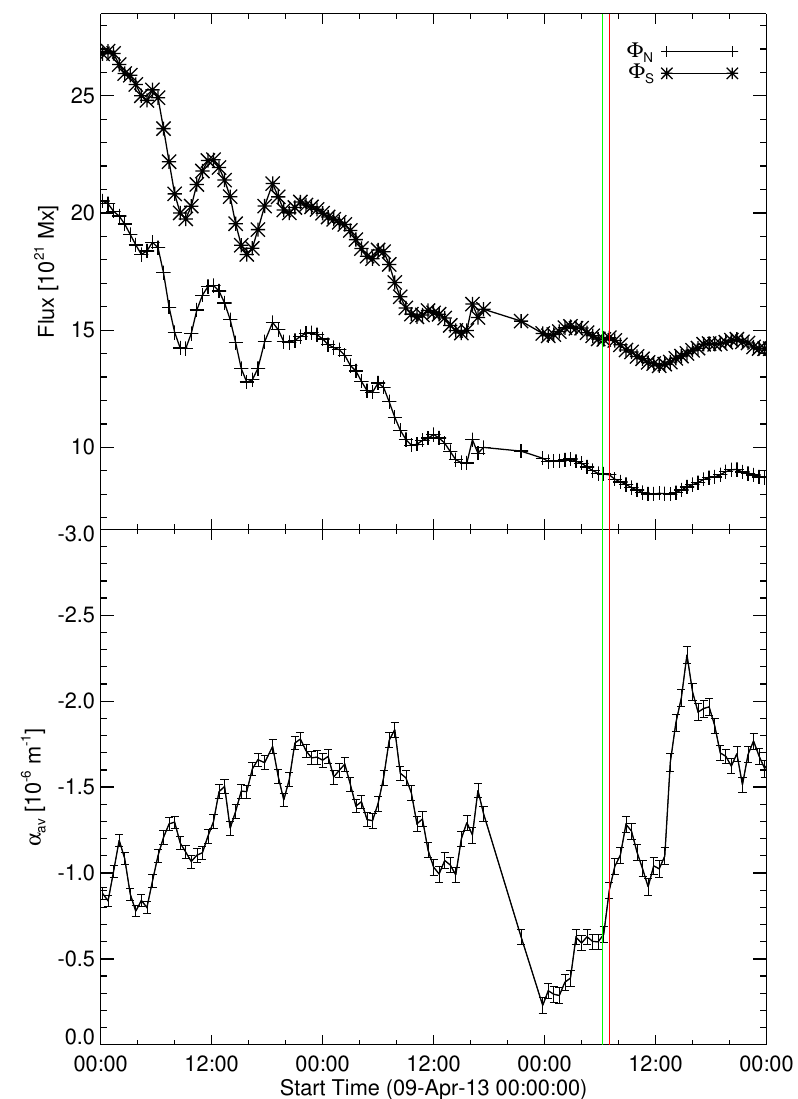}
\caption{{\bf top panel:} The evolution of flux content in the entire AR 11719. The estimated net line-of-sight flux from north and south magnetic polarities with respect to time is plotted, respectively. The monotonic decrease of the flux in the both polarities infers to possible flux cancellation. Note the steep decrease of the net flux till 15:00 UT on 10 April 2013 by which time the magnetic system of sigmoidal MFR formed (see Figure~\ref{Fig2} and~\ref{Fig4}). {\bf bottom panel:} Average alpha ($\alpha_{\rm av}$) evaluated over the entire active region with respect to time for three days. Vertical green and red lines indicate onset times of fluxrope rise and M6.5 flare, respectively.}
\label{Fig5}
\end{figure}

The corresponding details of the evolution of magnetic fields at the photospheric surface are studied (See Figure~\ref{Fig4}) by vector magnetic field observations. As explained earlier, the magnetic flux regions show little evolution during three days with respect to their motion. The time evolution of area integrated flux shows details about the change in the flux content. Indeed, the time profile of absolute flux in the active region (Figure~\ref{Fig5}, top panel) indicates a continuous decrease of magnetic flux from April 9 till the eruption onset on April 11. During this period, the north polarity flux (south polarity) decreased by 56\% (48\%) from its initial value at the early hours on 9 April. This decay or decrease includes polarity regions having flux rope legs and the overlying loops. Note that these profiles have 12 hour periodic variations probably related to orbital effects of the HMI instrument which introduce variations in the field measurements both spatially and temporally \citep{hoeksema2014}. 

As the time duration is about two days over which this significant decrease of the fluxes observed, it is likely related to general evolution of the active region in addition to slow reconnection and implies the usual scenario in decaying active regions. Further, to identify the size distribution of these fluxes, we overlaid contours of $\pm120$G. From snapshots of three successive days, all identified magnetic regions evolve with decreasing spatial distribution of flux. Since one of the legs of the MFR is associated with positive flux region N1 and N2, we isolate them (yellow contour) and measured the flux content at different snapshots (Figure~\ref{Fig2}). Consistent with the smearing flux distribution within this contour region, the measured net positive flux (See Table~\ref{tab1}) decreases from $4.27\times10^{21}$ Mx on April 9 to a significant value of $1.98\times10^ {21}$Mx on April 11, 2013. A similar measure of the main sunspot flux from which the MFR originated implies the same conclusion. 

These evolving conditions of fluxes support the sustainability and further development of the MFR. Because the MFR channel is already formed well before the observation on 9 April, under the slow motions of the flux regions that are mostly driven by granular motions, reconnection of the field lines will reinforce the MFR by added axial flux. Field lines connecting sunspot S1 to N1, possibly reconnect in a traditional tether cutting fashion \citep{moore2001} at the polarity inversion line with those connecting S2 and N2. However, those branches of field lines overlying the MFR from S2 to diffused patch of N2 provides downward tension force to balance the upward force of the MFR due to enhanced currents developed by the continuously added axial flux. For the occurrence of the successful eruption, the force balance should lost either by the weakening the overlying flux or by strengthening the axial flux in the MFR. Based on the observations, we argue that both of them could be conducive factors. As described earlier, the overlying flux bundle exhibits kink rise motion, which might allow for the initiation. The kink rise motion of the flux bundle signifies the increasing twist, which means enhanced currents and more axial flux in the MFR, supporting the second argument for force imbalance. In the absence of accurate twist measurements at the fluxrope leg, the first argument is the most reasonable to consider under the action play. Looking at this point, parametric and numerical studies \citep{aulanier2010, amari2010} suggest a limiting value of the axial flux of the rope (6-10\%) in the active region flux, beyond which force balance be lost. In decaying active regions, this value is in the range of 10-14\% \citep{bobra2008, suy2009, savcheva2009}. In our case, although it is difficult to separate these fluxes, for a quantitative approximation, the axial flux in the contour region (Table~\ref{tab1}) is calculated in a consistent range (9-14\%) of the total active region flux manifesting overlying polaidal flux. So, these results delineate the marginal stability conditions for the MFR.  

In order to further support the scenario of the formation or development of flux rope, we compute the parameter 

\begin{equation}
{{\alpha }_{av}}=\frac{\sum{{{J}_{z}}(x,y) sign[{{B}_{z}}(x,y)]}}{\sum{|{{B}_{z}}|}}
\end{equation}
as a proxy for twist of ARs. It is derived from the assumption of force-free magnetic fields and signifies the extent of twist of field lines due to field aligned currents. The error in $\alpha_{\rm av}$ is deduced from the least squared regression in the plot of vertical field $B_z$ and vertical current $J_z$ and is given by
\begin{equation}
\delta \alpha _{av}^{2}=\frac{{{\sum{\left[ {{J}_{z}}(x,y)-{{\alpha }_{av}}{{B}_{z}}(x,y) \right]}}^{2}}/|{{B}_{z}}(x,y)|}{(N-1)\sum{|{{B}_{z}}(x,y)}|}
\end{equation}
Where N is the number of pixels with $|Bt|>120$G (lower cut-off for the transverse field). Its time evolution is plotted in Figure~\ref{Fig3}. The time of rise motion of the flux rope and its eruption are indicate by vertical (green, red) lines. The error limit is much smaller than the mean value. Note the negative sign of $\alpha_{av}$ implies left handed helicity consistent with that of the inverse S-sigmoid. The twist parameter has no corresponding evolution with the decreasing flux till 11 April. It increased on 9 April following a gradual decrease on 10 April. Then it commences increasing from 11 April till 15:00UT to a value of -$2.3\times10^{-8}m^{-1}$. Together with decreasing flux content in the AR, the increasing average twist prominently from 11 April, characterize the conditions to augment the existing sigmoidal structure as inferred from EVU observations.

\begin{figure}[h!]
\centering
\includegraphics[width=.4\textwidth]{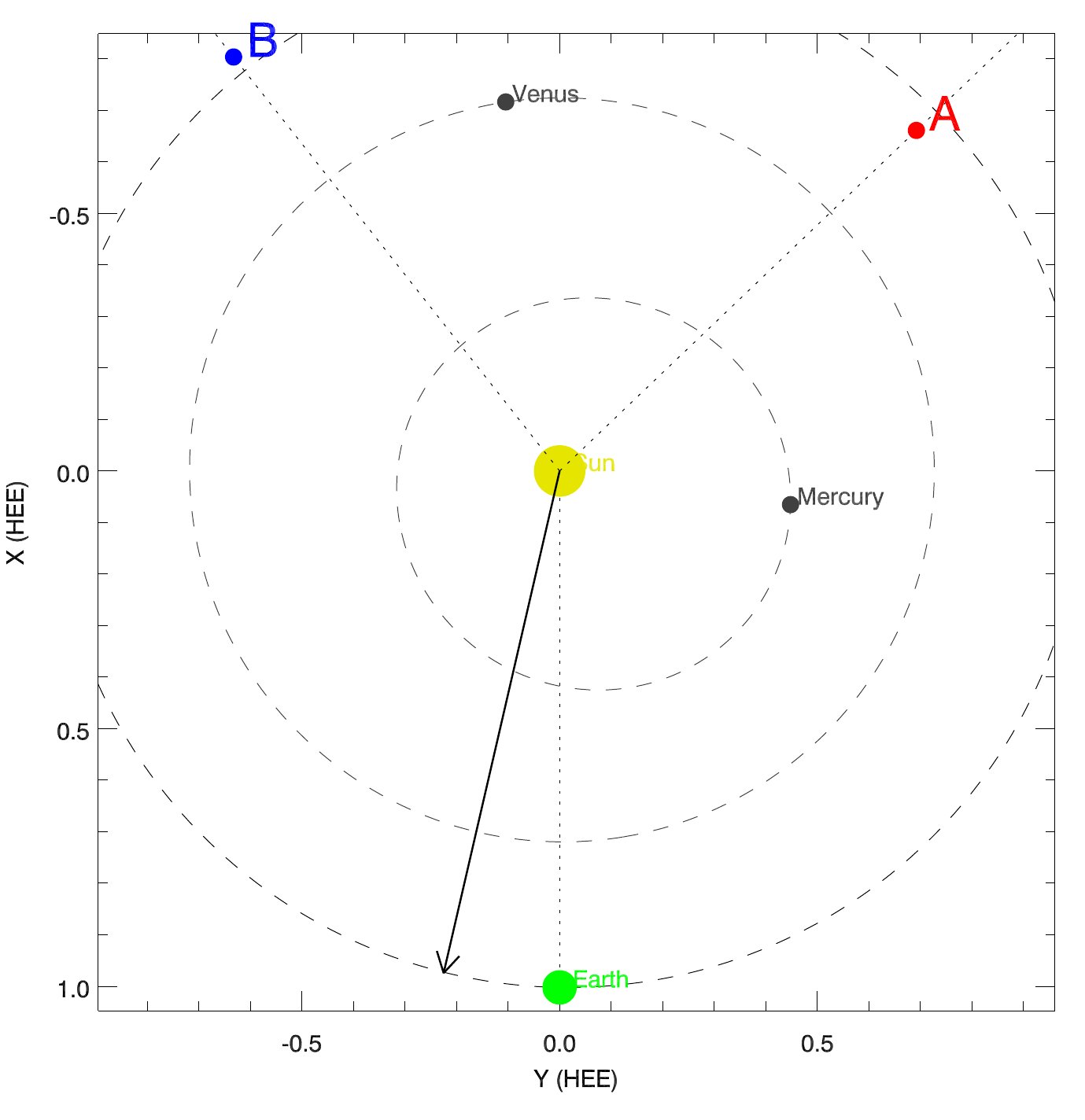}
\caption{The solar system upto 1 AU in the ecliptic plane. On April 11, 2013, the STEREO-A and B are separated by $276^o$ and located $133^o$ West and $141^o$ East, respectively with respect to Sun-Earth line. Arrowed line shows a possible direction ($13^o$ away from Sun-Earth line) of CME propagation towards the Earth. The coordinates are heliocentric Earth ecliptic Cartesian coordinates in units of 1AU. }
\label{Fig6}
\end{figure}

\section{The Eruption Mechanism}
\label{sec5}
At the time of eruption, the source AR contains a filament as seen in cool temperatures of AIA/304\AA, and over which a main inverse S-shaped sigmoid, and multiple twisted flux bundles visible in AIA hot pass bands mainly 131, 94\AA [last panel in Figure~\ref{Fig2} and~\ref{Fig3}]. Interpreting this sigmoid system in terms of MFR models, the initiation mechanism together with morphological, thermal, kinetic and magnetic properties of this eruption event was studied in great detail by VZ14. This sigmoidal MFR starts with a slow rise from 06:15UT followed by fast rise from 06:40UT and eventual eruption. Consistent with our observational signatures of development of the MFR over days, the results of their study suggests that the cancelling fluxes are prime factors to the monotonous twisting of the FR system reaching to a critical state to trigger kink instability. The support for this instability comes from the observed kink-like evolution of the overlying flux bundle (Figure~\ref{Fig3} and Figure 2 in VZ14) and the co-spatial localized distribution of increased twist parameter in the main sunspot from where the MFR originated. This instability likely initiates the rise motion until a critical height for a possible onset of torus instability, from which the MFR likely undergoes self similar expansion and outward propagation, subsequently resulting in the eventual eruption of Earth-directed CME. Note the analysis based on MFR insertion model (Table~\ref{tab1}), the net positive flux (as a proxy for axial flux) over time  indicates marginal stability to this MFR system. The preceding flare of GOES M6.5/3B class (from 06:55UT, see Figure~\ref{Fig1}d) is a consequence of progressive reconnection in thinning current sheet underneath the rising FR. 

\section{Orientation of the CME MFR in Extended Corona}
\label{sec6}
\begin{figure*}[htb!]
\centering
\includegraphics[width=.98\textwidth]{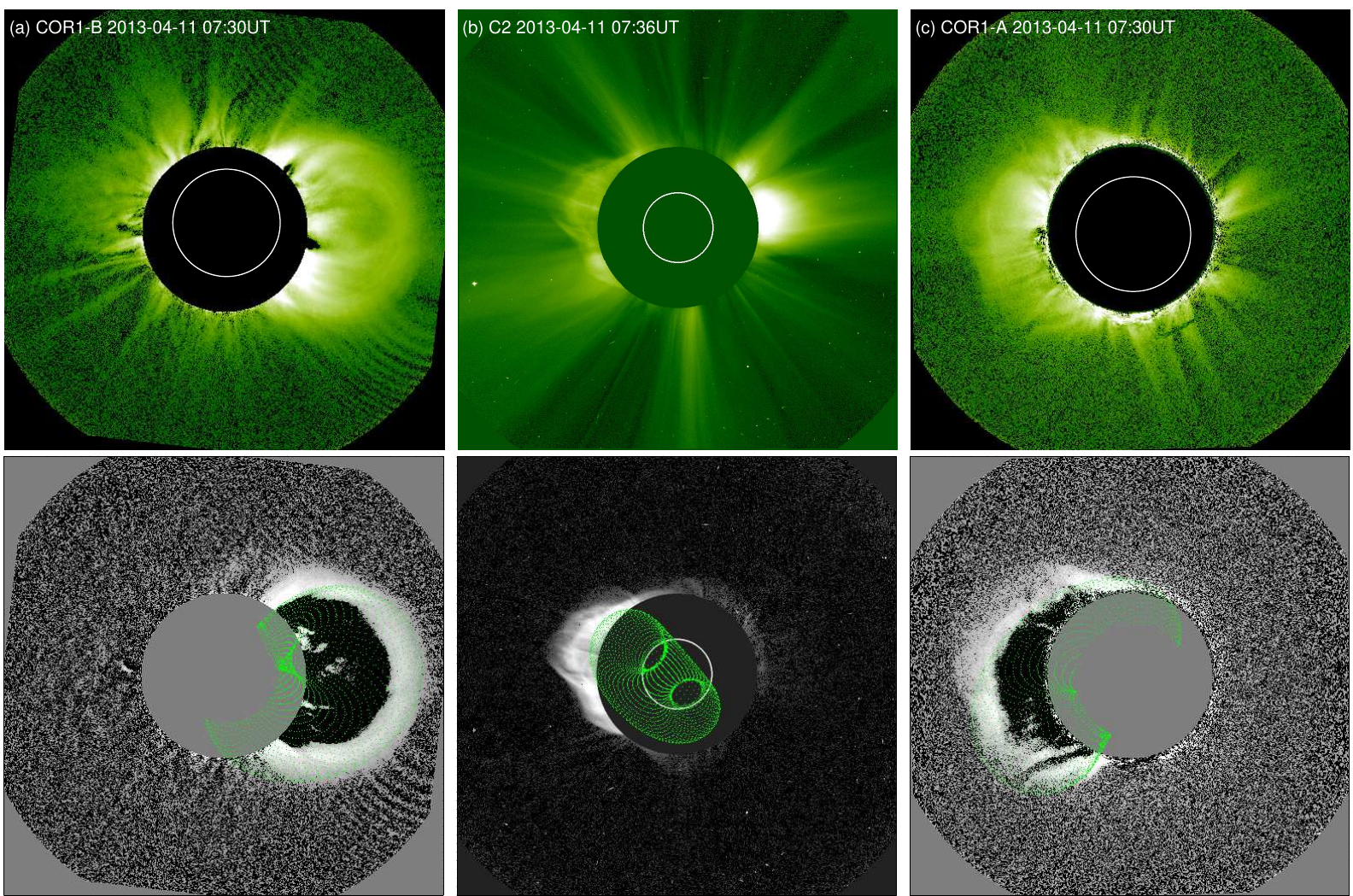}
\caption{ Top panels: CME morphology observed from COR1-A, C2, and COR1-B respectively at 07:30UT on April 11 2013. Bottom panels: Overplot of the best fitted wireframe of MFR in GCS model. Note that in C2 the CME is halo directed toward Earth, whereas it directs away from COR-A on farside.  Note the GCS parameters $\gamma=-45^o$, $\alpha=50^o$, $\kappa=0.45$ and $h=4R_{\odot}$ reproduce this CME morphology. }
\label{Fig7}
\end{figure*}

\begin{figure*}[htb!]
\centering
\includegraphics[width=.98\textwidth]{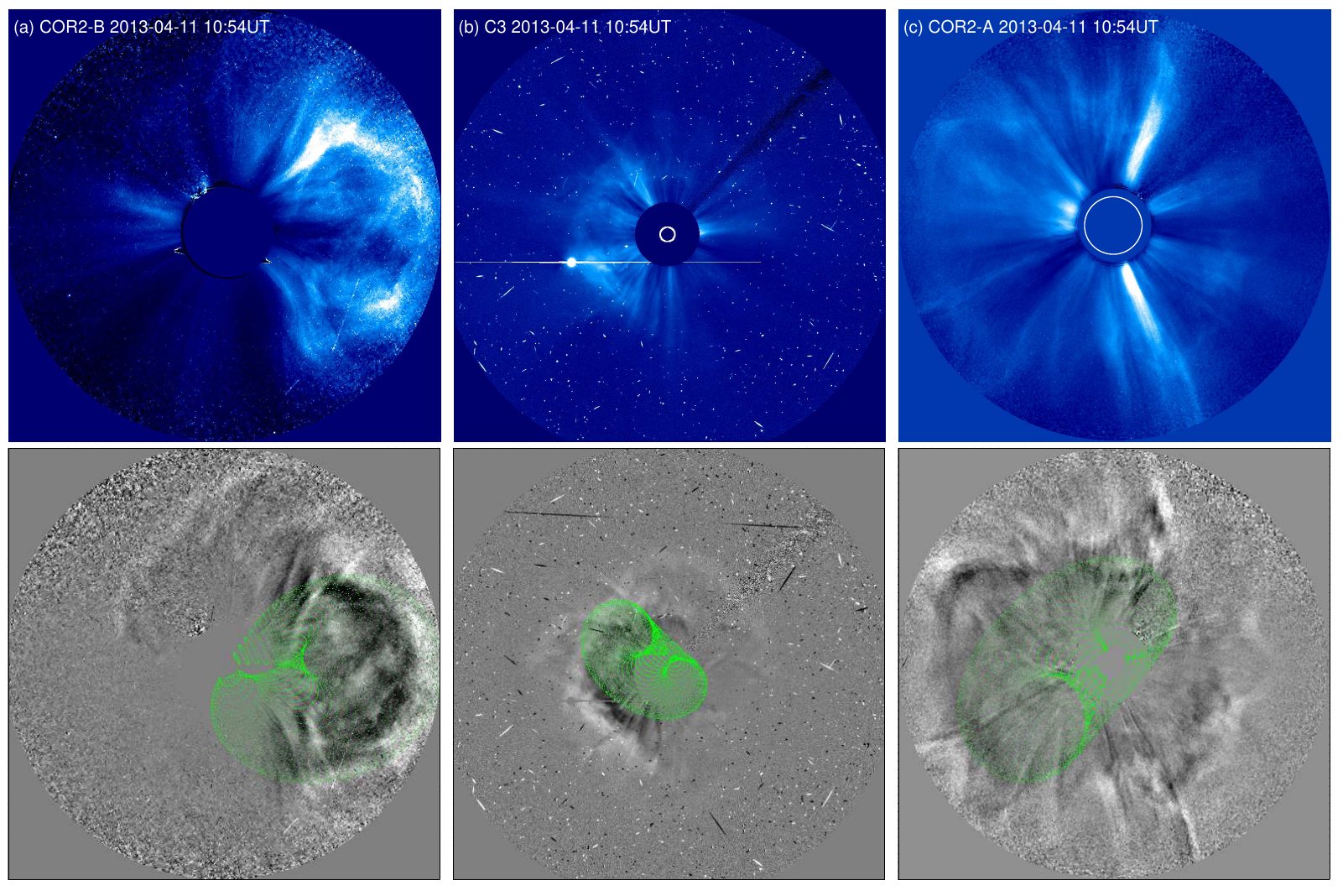}
\caption{ Top panels: CME morphology observed from COR2-A, C3, and COR2-B respectively at 10:54UT on April 11, 2013. Bottom panels: Overplot of the best fitted wireframe of MFR in GCS model. Note that in C3 the CME is halo directed toward Earth, whereas it directs away from COR-A on farside.  Note the GCS parameters $\gamma=-45^o$, $\alpha=50^o$, $\kappa=0.45$ and $h=17.5R_{\odot}$ reproduce this CME morphology. }
\label{Fig8}
\end{figure*}

During the launch of 2013 April 11 CME from the Sun, the solar system in the ecliptic plane upto 1 AU and the positions of STEREO-A and B spacecrafts are depicted in the plot of Figure~\ref{Fig6}. The eruption is seen on the solar disc from the Earth (SDO, LASCO) perspective and the CME was observed by SOHO/LASCO-C2 at 07:24 UT onwards as a full halo CME with a speed of the CME\footnote{\url{http://cdaw.gsfc.nasa.gov/CME_list/UNIVERSAL/2013_04/}} in LASCO FOV was 861 km/s  and was found to be decelerating. STEREO-A was located $133^o$ West and STEREO-B was $141^o$ East of the Earth, and so they could not observe the CME MFR during initial times within the low corona. However, the separation angle between them ($276^o$) makes it possible to observe the MFR as a bulk structure of the CME that emerged out of the occulter after 07:10 UT in the East limb of STEREO-A and after 07:30UT in the West limb of STEREO-B on April 11 (Figure~\ref{Fig7} and~\ref{Fig8}, top panels). The MFR structure (as seen in classical three part structured CME) could be identified as a dark cavity in the CME morphology in COR1-B.  Since this CME is halo event from COR-A and LASCO views, its three part structure could not be identified there. The new definitions suggest that the CMEs which travel beyond $10R_{\odot}$ are being supported by MFR topology and ruling out other possibilities of loops and jets \citep{vourlidas2013}. Therefore, our description of MFR topology to the observed morphology of source region sigmoid (sections~\ref{sec3}-\ref{sec5}) and the CME fits well with the observed properties in both the low and upper corona as well. 

In order to reveal the underlying large scale structure and orientation of MFR by exploiting the plane of sky projected images of plasma structures, we need to determine its three-dimensional reconstruction using simultaneous imaging observations from multiple view points. For this purpose, we employed simultaneous three-point i.e., the SECCHI/COR-B, LASCO-C2/C3 and SECCHI/COR-A white light observations of the CME, with a model called Graduated Cylindrical Shell (GCS; \citealt{thernisien2009}. In this model, the large scale structure of MFR is approximated by two shapes; the conical legs and the curved (tubular) fronts, which is known as the "hollow croissant". The involved inherent assumption in this model is that the magnetic orientation of an erupting CME MFR is constrained by the pre-eruptive magnetic configuration in the source AR, invoked after a study of many CMEs and their source region magnetic configurations \citep{cremades2004}. Before applying this model, a suitable background image was subtracted from a sequence of processed total brightness images.  

\begin{table}[h!]
\caption{GCS fitting parameters at different times}
\begin{tabular}{llll}
\hline
Time [UT], April 11 	&	 $\kappa$ & $\alpha$ [deg] &	height [$R_{\odot}$] \\
\hline	
07:30   											&	0.45			& 50								&  4.0			\\
08:24  											& 0.45				& 50								& 8.5			\\
09:54 												&0.45				& 50								&13.5		\\
10:54 												& 0.45				& 50								&17.5		\\
\hline
		\end{tabular}
	\label{tab2}
\end{table}

Figure~\ref{Fig7} and~\ref{Fig8} (bottom panels) show the difference images of the CME at 07:30-07:24UT [COR1-A, C2 and COR1-B] and 10:54-10:24 UT [COR2-A, C3 and COR2-B] on 11 April, respectively. These panels are overlaid with a wireframe (green) of the MFR retrieved from an approximated fit of GCS model. While fitting the GCS model, we input the source location position (latitude of $9^o$ North, longitude of $13^o$ East), tilt-angle ($\gamma=-45^o$) of magnetic polarity inversion line (see Figure~\ref{Fig2}) and vary the height, aspect-ratio ($\kappa$, major and minor radius of the MFR), half-angle ($\alpha$) until achieving the best visual fit to the observed Thomson scattered bright emission of the CME. Table~\ref{tab2} summarizes these fitted parameters at four different epochs of CME propagation in the inner corona. At all four time instants, an half angle of $50^o$ and aspect ratio of 0.45 gives a best resemblance to the projected MFR on to the plane of the sky. We point that the constant value of $\kappa$ during the outward motion of MFR indicates its self-similar nature of expansion as observed for many CME events \citep{subramanian2014} in the inner corona. Given radial positions at successive time instances, intermediate velocities and their average value can be estimated. Thus calculated instantaneous velocities 967km/s [07:30-08:24UT], 644km/s [08:24-09:54UT], 773km/s [09:54-10:54UT] arrive to an average value of 795km/s. Note the published average velocity in C2 fov from plane of sky observations is 860km/s. From this analysis of Earthward-directed CME MFR, it is clear that the determination of the parameters defining the MFR orientation from three-point observations is somewhat a near approximation and useful to reproduce oblique cases when MFR is not an edge-on (limb) event. As MFRs rotate on their outward propagation \citep{lynch2009} due to their inherent twist, it is quite possible that this MFR tilt be $-55^o$ within COR2 FOV as this value too fits the observed morphology very well.

\section{Propagation of the CME MFR in the Heliosphere}
\label{sec7}
\subsection{Kinematics from Stereoscopic and Single Spacecraft Reconstruction Methods}
\begin{figure*}[htb!]
\centering
\includegraphics[width=.98\textwidth]{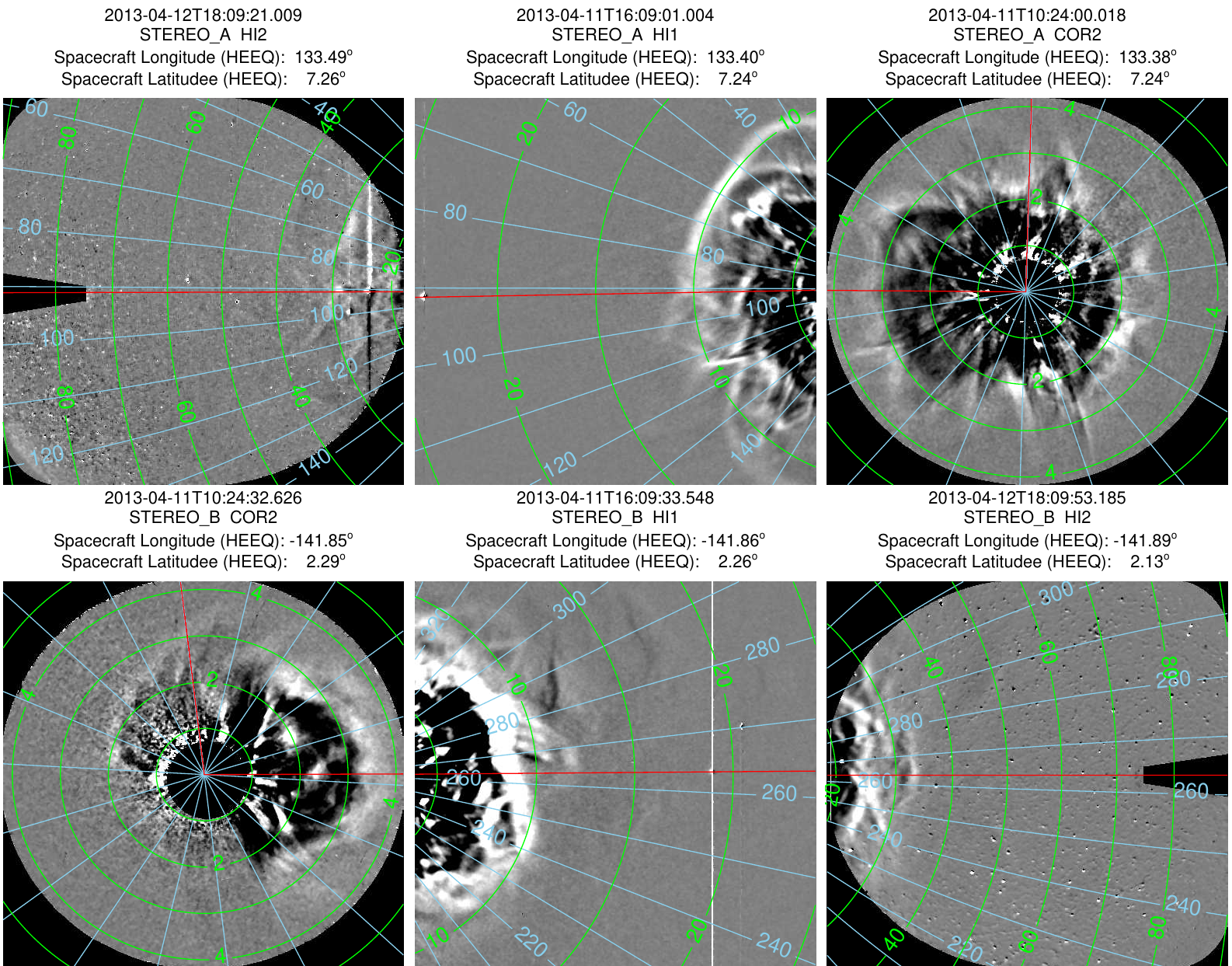}
\caption{{\bf top row:} Left, middle, and right panels show the evolution of 2013 April 11 CME in the running difference image of HI2-A, HI1-A, and COR2-A respectively. {\bf bottom row}:  Running difference images from COR2-B, HI1-B, HI2-B in Left, middle and right panels, respectively. The contours of the elongation angle (green) and the position angle (blue) are also deployed on each image. In all panels, horizontal red line is along the ecliptic and indicates the position angle of the Earth and vertical red line (left panel)  refers $0^{o}$ position angle.}
\label{Fig9}
\end{figure*}

\begin{figure}[h!]
\centering
\includegraphics[width=.48\textwidth]{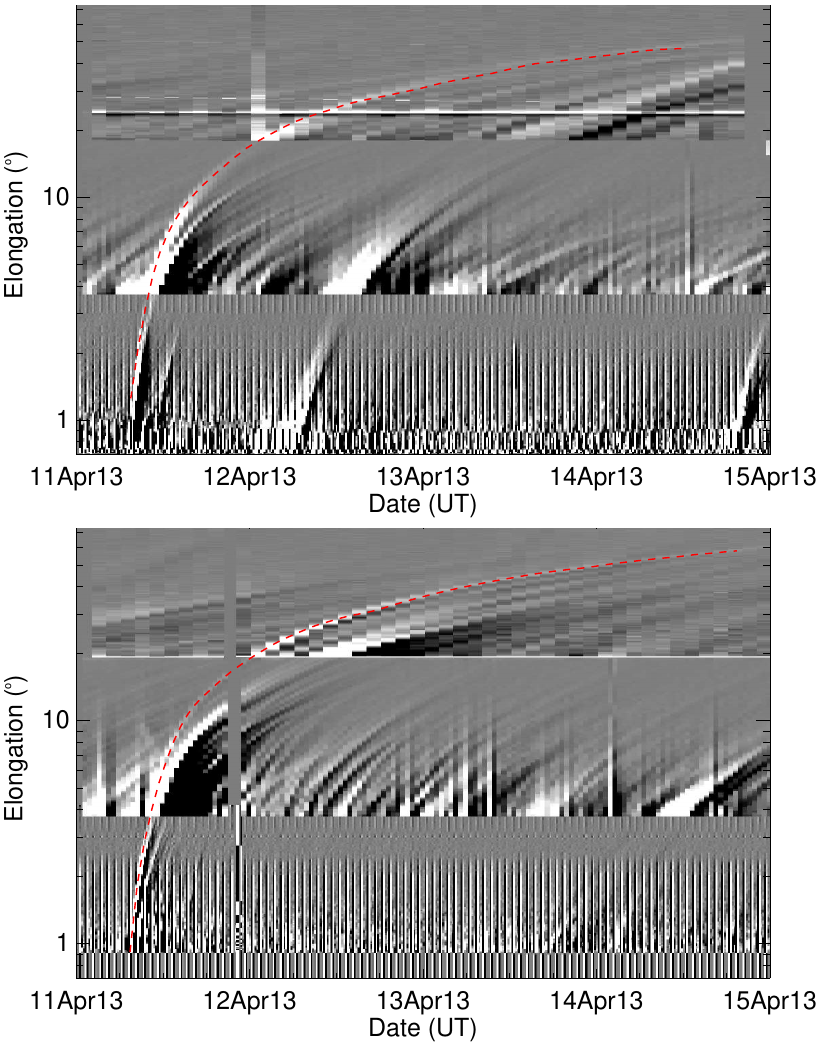}
\caption{Time-elongation maps (J-map) in the ecliptic plane for spacecrafts A(top) and B (bottom) on STEREO using running difference images of COR2, HI1 and HI2. The tracks of the 2013 April 11 CME is indicated with a red curve.}
\label{Fig10}
\end{figure}

It has been shown that different parts (structures) of the CME propagate and reach the Earth in the same order as observed in remote observations \citep{liuying2010, byrne2010, howard2012}. We acknowledge the difficulty in tracking the MFR of a CME using running difference images from HI. However, tracking of density enhanced feature as leading edge of the CME, which is solar wind or coronal plasma pile-up ahead of the MFR and identified as shock-sheath region in in situ data, is conveniently possible \citep{davies2009, mostl2010, mishra2013}.  Assuming that the 3D kinematics of CME leading edge will mimic that of the associated MFR too, we have estimated the 3D kinematics of the leading edge using COR2, HI1 and HI2 observations from STEREO (Figure~\ref{Fig9}). From them, time-elongation maps (J-map; \citealt{davies2009}) are constructed as shown in Figure~\ref{Fig10}. Positive inclination bright curves in the J-maps indicate the motion of the CME features away from the Sun. Using these J-maps, we tracked the bright leading front manually and derived their elongation-time profiles (dashed red line). 

The three most widely used stereoscopic reconstruction methods to estimate the kinematics of CMEs using SECCHI/HI observations are the Geometric Triangulation (GT; \citealt{liuying2010}), Tangent to a Sphere (TAS; \citealt{lugaz2010a}) method and stereoscopic self-similar expansion (SSSE; \citealt{davies2013}) method. These methods differ by assuming a different geometry for the CME. \citep{davies2013} showed that GT and TAS methods are the special cases of SSSE method corresponding to cross-sectional angular half-width ($\lambda$) of the CME as $0^o$ and $90^o$, respectively. We therefore, implemented the SSSE method to estimate the kinematics for the present case. 

Using the elongation measurements (Figure~\ref{Fig10}) and position of STEREO spacecraft, the estimated kinematics from SSSE method for different values of $\lambda$ is shown in Figure 11. The plots for $\lambda = 0, 30, 60^o$ are shown till the CME reaches L1 location, i.e. approx. $214R_{\odot}$ from the Sun. This is done to examine the role of cross-sectional angular extent of the CME in estimating the kinematics. The estimated kinematics with $\lambda$ equal to $90^o$ is shown for relatively larger distances. It is evident that the estimated kinematics differs for different $\lambda$. It is to be noted that although distances and speeds derived from these methods vary significantly, the estimates of direction and its trend from these methods are within $10^o$. Estimates of direction from all the methods suggest that the CME is propagating towards the East of the Sun-Earth line in the ecliptic plane. This is consistent with the solar source location of the CME and the estimated direction in COR FOV from the GCS model. 

\begin{figure*}[htb!]
\centering 
\includegraphics[width=.75\textwidth]{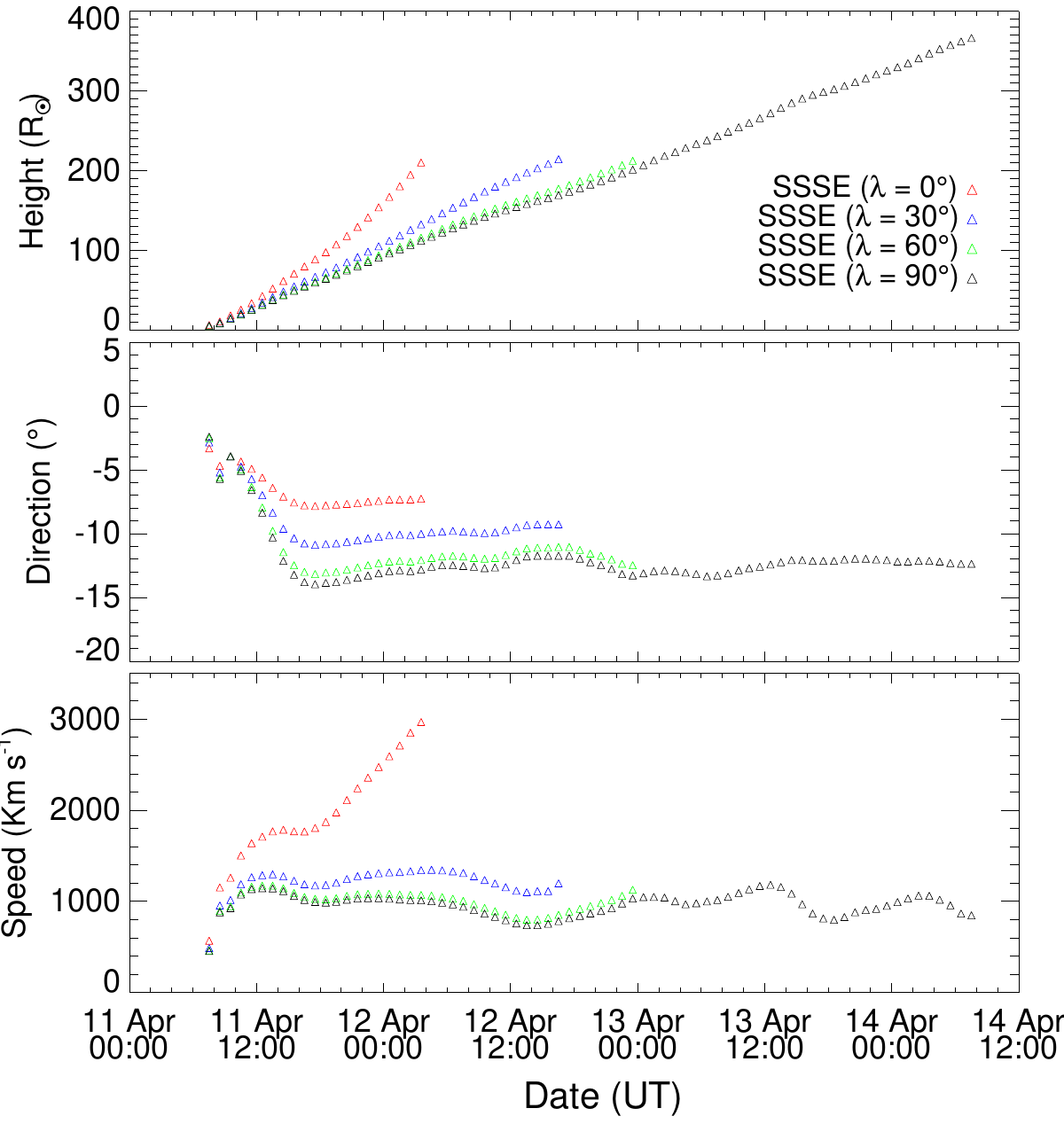}
\caption{From top to bottom, the panels show the distance, propagation direction, and speed, respectively, of the 2013 April 11 CME MFR. The kinematics shown with red, blue, green and black are derived from SSSE method with $\lambda$ equal to 0, 30, 60 and $90^o$, respectively. The estimated speed has systematic fluctuations due to numerical differentiation and do not represent the real effect on the CME. It highlights the propagation of error into the kinematics because of even a little uncertain tracking of the CME. Such a procedure for estimating the speed has potential to reveal crucial short time variations in speed. However, the smoothed speed obtained by fitting the estimated distance-time curve into a polynomial may hide the short time variations in the CME speed.}
\label{Fig11}
\end{figure*}

We admit that there are several sources of errors (geometry, Thomson scattering, optically thin nature of CMEs, breakdown of assumptions considered in the methods itself) in implementing these methods and quantification of such errors is extremely difficult. However, to examine the effect of uncertainties in the estimated kinematics, we considered an error of 5 pixels in the measurements of elongation angle. Such an error corresponds to uncertainties of 0.02 degree, 0.1 degree and 0.35 degree in the derived elongation angles in COR2, HI1 and HI2 FOV, respectively. Corresponding to these elongation uncertainties, the calculated error for SSSE (with $\lambda=0$) method maximally reaches up to $0.5R_{\odot}$ in distance, less than few degree in direction and few tens of km/s for speed. These errors certainly seem to be small but of course, they do not reflect the total error in derived kinematics. Further study is required to quantify the actual errors because of several invalid idealistic assumptions in the methods.

Also, at higher elongation (greater than approx. $40R_{\odot}$), larger variations in the estimated kinematics from SSSE method for different $\lambda$ are noted. From Figure~\ref{Fig11}, it is clear that the estimated distance from GT method increased around 18:00 UT on April 11 at $90R_{\odot}$ from the Sun. Such an unphysically fast increase in distance and speed is meaningless due to the absence of forces capable of accounting for this acceleration at distances farther from the Sun \citep{cargill2004, vrsnak2010}. This is more likely because of improper use of SSSE with $\lambda$ equal to $0^o$ (i.e. GT) method, especially for this CME which is propagating away from the observer. This late acceleration is significantly reduced (still unphysical) if a higher value of $\lambda$ is considered. SSSE with $\lambda$ equal to $90^o$ (i.e. TAS) method gives the lowest limit of the estimated distance and speed values. The kinematics from the TAS method is calculated up to $365R_{\odot}$ from the Sun.

The aforementioned facts highlight that the assumptions made in the GT method are not valid for a CME propagating away from the observer. Also, the spherical front approximation in SSSE method with $\lambda$ equal to nonzero value, becomes worse due to flattening of the CME front on its interaction with solar wind. These limitations start to play a crucial role much nearer to the Sun for a far-sided CME than for front sided CME \citep{liuying2013}. However, the distance beyond which these effects are crucial depends on the direction of propagation of the CME and its size. We note that for a CME propagating at larger angle from the Sun-observer line, its kinematics depend more on the chosen value of $\lambda$. This is because the flanks (not nose) for such CMEs are always observed from the observer and even a little change in its radius of curvature leads to huge difference in the estimated kinematics from the SSSE method. It is noted that if direction of propagation of a CME from sun-observer line becomes more than $90^o$, an unphysical acceleration will be estimated from the SSSE methods irrespective of chosen $\lambda$ value. However, such an acceleration is reduced with higher value of $\lambda$ and gives more accurate kinematics. 

To assess the relative performance, we also applied single spacecraft reconstruction methods on STEREO-A and B observations. We employed the Fixed-Phi (FP: \citealt{kahler2007}), Harmonic Mean (HM: \citealt{lugaz2009}), Self-Similar Expansion (SSE: \citealt{davies2012}) methods on the derived time-elongation variations of the CME (Figure~\ref{Fig10}). These methods essentially convert the elongation into distance from the Sun assuming a fixed direction (longitude, here $13^o$ East) of CME propagation as an input. 

On applying the above methods on the STEREO-A and B observations, we noticed that estimated kinematics also largely overestimates the speed (approx. 900-1200 km/s even beyond $100R_{\odot}$) of the CME and the method becomes completely unreliable once the CME reaches higher elongations. The direction of propagation of CMEs is $146^o$ and $128^o$ away from the line connecting the Sun with STEREO-A and B spacecraft, respectively. As the CME propagating in Eastward is little closer to $90^o$ from the STEREO-B spacecraft, the derived kinematics and arrival time from this spacecraft are more accurate than using STEREO-A  observations. Among all the three single spacecraft methods, the most inaccurate results are obtained from the FP method and less inaccurate are from the HM method. This probably confirms the assumption that the larger structure of CME is somewhat suitable for estimating the time varying profile of the CME kinematics. However, the failure of these single spacecraft methods at higher elongation could be due to real deflection or artificial deflection because of expansion or/and due to changes in the approximated idealized structure \citep{wood2010, howard2011, mishra2014a}. 

We further applied the fitting version of the three single spacecraft reconstruction methods, namely Fixed-Phi Fitting (FPF: \citealt{sheeley2008}), Harmonic Mean Fitting (HMF: \citealt{mostl2011}), and Self-Similar Expansion Fitting (SSEF: \citealt{davies2012}). Noting the expressions for the elongation as a function of speed and direction from the earlier FP, HM and SSE methods, these methods fit the observed elongation-time profile of the CME to an analytical function. From the FPF method on STEREO-A, we found that CME speed as 654 km/s, propagation direction as $99^o$ from STEREO-A spacecraft and its launch time at 11 April 05:05. Similarly, we also applied the HMF and SSEF methods (with $\lambda=50^o$) and obtained the speed, propagation direction and launch time of the CME. The results obtained for STEREO-A and B are shown in Figure~\ref{Fig13}~and~\ref{Fig14}, respectively and summarized in Table~\ref{tab3}. In these figures, the results for SSEF, which fall between the FPF and HMF methods, are not plotted to avoid cluttering.  

\begin{figure}[h!]
\centering
\includegraphics[width=.49\textwidth]{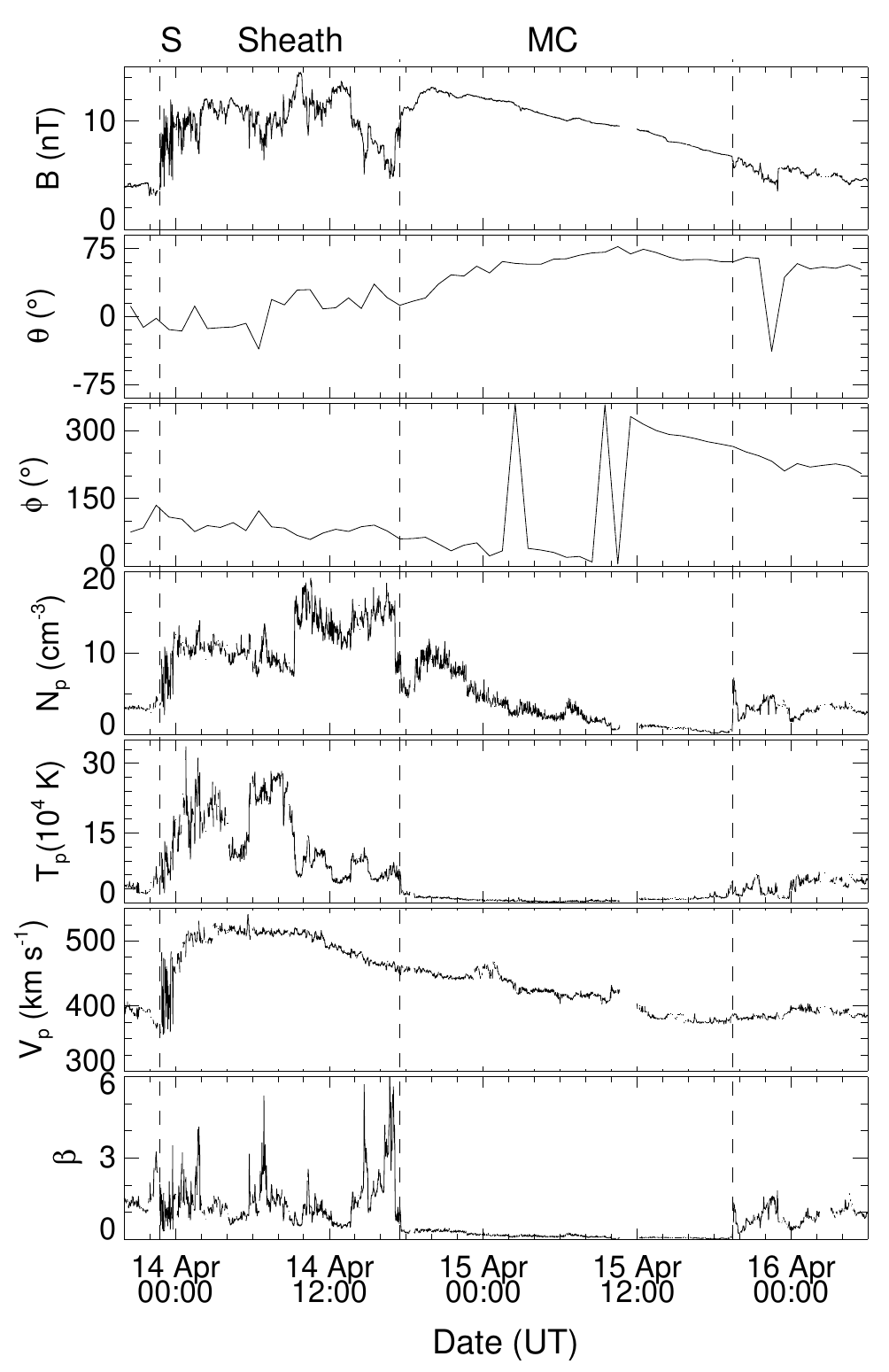}
\caption{From top to bottom, panels show the magnitude of magnetic field, latitude, longitude of magnetic field vector, density, temperature, speed, and plasma $\beta$ with respect to time. From the left, the first, second and third vertical lines mark the arrival of CME driven shock (at 22:50UT on 13th), leading (at 17:35UT on 14th) and trailing edge (19:20UT on 15th) of magnetic cloud, respectively.}\label{Fig12}
\end{figure}

\begin{figure}[h!]
\centering
\includegraphics[width=.49\textwidth]{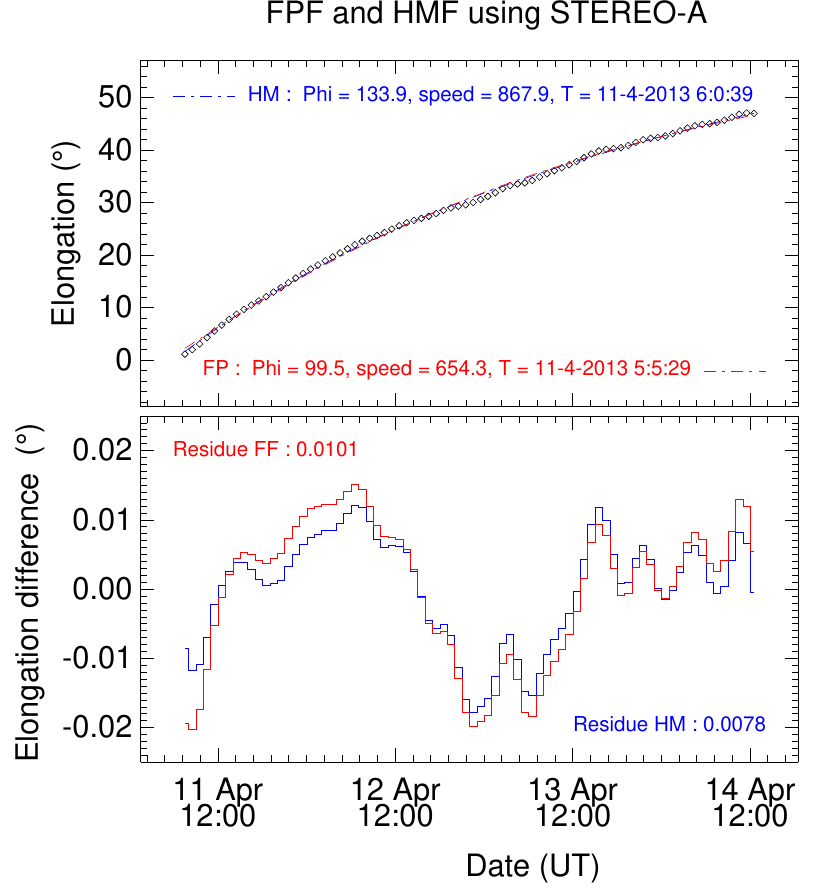}
\caption{Best fit FPF and HMF results, for the tracked CME feature from STEREO-A location, are shown with red and blue colors. {\bf top:} time-elongation variation theoretically derived from FPF and HMF approaches are overplotted on the observed elongation-time variation of the tracked feature. {\bf bottom:} Residuals between the derived elongation from fitting functions and the observations. In the top panel, the labels Phi, speed and T are the approximated constant speed, direction of propagation from the STEREO-A, and launch time of the CME on the Sun, respectively.}\label{Fig13}
\end{figure}

\begin{figure}[h!]
\centering
\includegraphics[width=.49\textwidth]{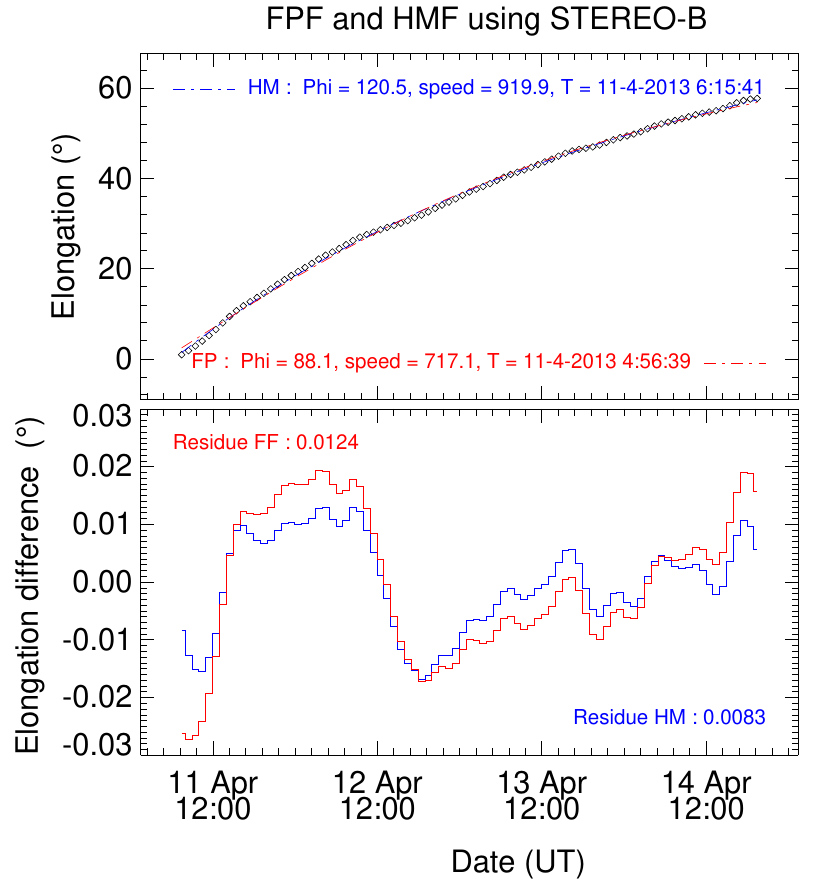}
\caption{Best fit FPF and HMF results, for the tracked CME feature from STEREO-B location, are shown with red and blue colors. {\bf top:} time-elongation variation theoretically derived from FPF and HMF approaches are overplotted on the observed elongation-time variation of the tracked feature. {\bf bottom:} Residuals (i.e. the difference) between the derived elongation from fitting functions and the observations. Residuals are estimated in a similar procedure to \citet{mostl2011}. In the top panel, the labels Phi, speed and T are the approximated constant speed, direction of propagation from the STEREO-B, and launch time of CME on the Sun, respectively}\label{Fig14}
\end{figure}

Moreover, we have also analyzed the in-situ observations of this CME obtained from WIND spacecraft (Figure~\ref{Fig12}). The actual arrival time of the CME is marked by the arrival of a shock at 22:50 UT on April 13, which is followed by a prolonged (18 hr) sheath region. This sheath region is followed by a MC from 17:35UT on April 14, whose signatures are indicated by rotation of the magnetic field, an increase in the magnetic field strength to 14n and decrease in the temperature (order of $10^4$\,K). Duration of this MC traversal is approximately 25 hours at a speed of approximately 450 km/s, suggesting its width as approximately 0.27 AU, which is the average size of MCs as found by previous studies \citep{liuyrichardson2005,liuyrichardson2006, liuying2010}. The arrival of the shock (jump in density, temperature and speed) structure ahead of the CME is expected to match the estimated arrival of the tracked density enhanced feature in the heliosphere using the J-maps as shown in Figure~\ref{Fig10}. Therefore, for a quantitative analysis, we compared the estimated arrival time by various reconstruction methods to that of in-situ shock arrival and listed the errors in Table~\ref{Fig3}. 

The errors in the estimated arrival time from the SSSE method with $\lambda$ equal to 0, 30, 60, and $90^o$ are earlier than the actual time by 46, 31, 26 and 21.5 hr, respectively. This shows that the estimates of arrival time from HI observations have large errors due to overestimated speed values. In light of these findings, it appears that in such a case, SSSE methods with higher value of $\lambda$ (larger elongation from the Sun) provide better kinematics for the CME.

Based on the kinematics determined from the FP, HM and SSE methods, the arrival time of this CME is still estimated as 15-40hr before the actual arrival. The error in the kinematics may be because of inaccurate fixing of the propagation angle and therefore by changing the propagation angle of $\pm10^o$, we again repeated the analysis and still found the huge error in the arrival time of the CME. On applying  these methods, we inferred less erroneous kinematics obtained using the STEREO-B observations when the CME is fixed to propagate eastward from the Sun-Earth line and vice-versa for STEREO-A observations.
\begin{table*}[htb!]
\centering
\begin{threeparttable}
\caption{Summary of the propagation kinematics from Stereoscopic and Single Spacecraft reconstruction methods}
\begin{tabular}{llllll}
\hline
Method 	&		Speed\tnote{a} 								&  Direction\tnote{b} 									& Launch Time\tnote{c} 				& Arrival Time  		& Error in Arrival Time\tnote{d}  \\ 					      
                &   [km/s]                                 & [deg]                                  			&		[UT]                                    & [UT]                   &  [h] 				\\
\hline	
SSSE ($\lambda=0^o$)	& 2900  & -7				& 11 April 07:40       & 12 April 01:00     & -46			\\
SSSE ($\lambda=30^o$)& 1100  & -9 			& 11 April 07:40       & 12 April 16:05     & -31			\\
SSSE ($\lambda=60^o$)& 1140  & -12 		& 11 April 07:40       & 12 April 21:00     & -26				\\
SSSE ($\lambda=90^o$)& 1060  & -13     & 11 April 07:40      &  13 April 01:30      & -21.5				\\
FPF(ST-A)   			&  	654  				& 34					 &	11 April, 05:05			&	13 April, 19:55			& -3  		\\
FPF(ST-B)   			&  	717  				& -53					&	11 April, 04:56				&	13 April, 14:20			& -8.5     \\
SSEF(ST-A)   		&  	867  				& 0						&	11 April, 05:50				&	13 April, 05:15			& -17.5  \\
SSEF(ST-B)			  &  	919  				& -20					&	11 April, 06:01				&	13 April, 10:40			& -12.2  \\
HMF(ST-A)   			&  	979  				& -13					&	11 April, 06:00				&	13 April, 01:10			& -21.6	 \\
HMF(ST-B)   			&  	1016  			& -10					&	11 April, 06:15				&	13 April, 23:15			& -23.5  \\				
\hline
\end{tabular}
\begin{tablenotes}
\footnotesize
	\item[a] for SSSE methods, the quoted speed is transit value
	\item[b] direction with negative (positive) sign refers to CME propagation in eastward (westward) to the
Sun-Earth line.
\item[c] First COR2 appearance of leading edge is taken as launch time for SSSE methods
\item[d] negative (positive) value of error in arrival time refers earlier (late) arrival of the CME   
\end{tablenotes}
\label{tab3}
\end{threeparttable}  
\end{table*}    

We noted that the HMF and SSEF methods estimate the propagation speed of the CME apex, which is not exactly towards the Earth, therefore we have applied geometrical correction \citep{mostl2013} to estimate its speed in an off-apex direction. This geometrically corrected speed, which is less than its speed derived in the apex direction, is used to obtain the estimated arrival time of the CME at L1 point.

From Table~\ref{tab3}~it is obvious that, among all the single spacecraft reconstruction methods, the most accurate estimation of CME arrival time (within an error of 3-8 hr) is obtained by the FPF method and less accurate by the HMF method. However, if we believe the direction of propagation of CME estimated from GCS model as accurate, then estimation of the direction from FPF methods is erroneous in agreement with \citep{lugaz2010b}. We find that even the fitting methods overestimate the speed of the CME. In principle, these fitting methods always overestimate the speed of fast CME as they estimate a constant speed for CME and therefore they are unable to incorporate the real deceleration due to acting drag forces on the CME. This limitation of these methods is most obvious for CME propagating largely away from the Sun-Spacecraft line. If we bear the uncertainties with the direction of propagation of the CMEs, then FPF method does a good job for predicting arrival time of CME at L1. We infer that in the FPF method, the two sources of errors arising from not taking the geometry of the CME and physical deceleration into account actually cancels each other. However, any such physical deceleration is significantly counted in terms of geometrical (i.e. apparent) deceleration, if SSEF and HMF methods are used. The analysis of this event could help us to realize the potential differences in the results because of different fitting methods.

\begin{figure}[h!]
\centering
\includegraphics[width=.49\textwidth]{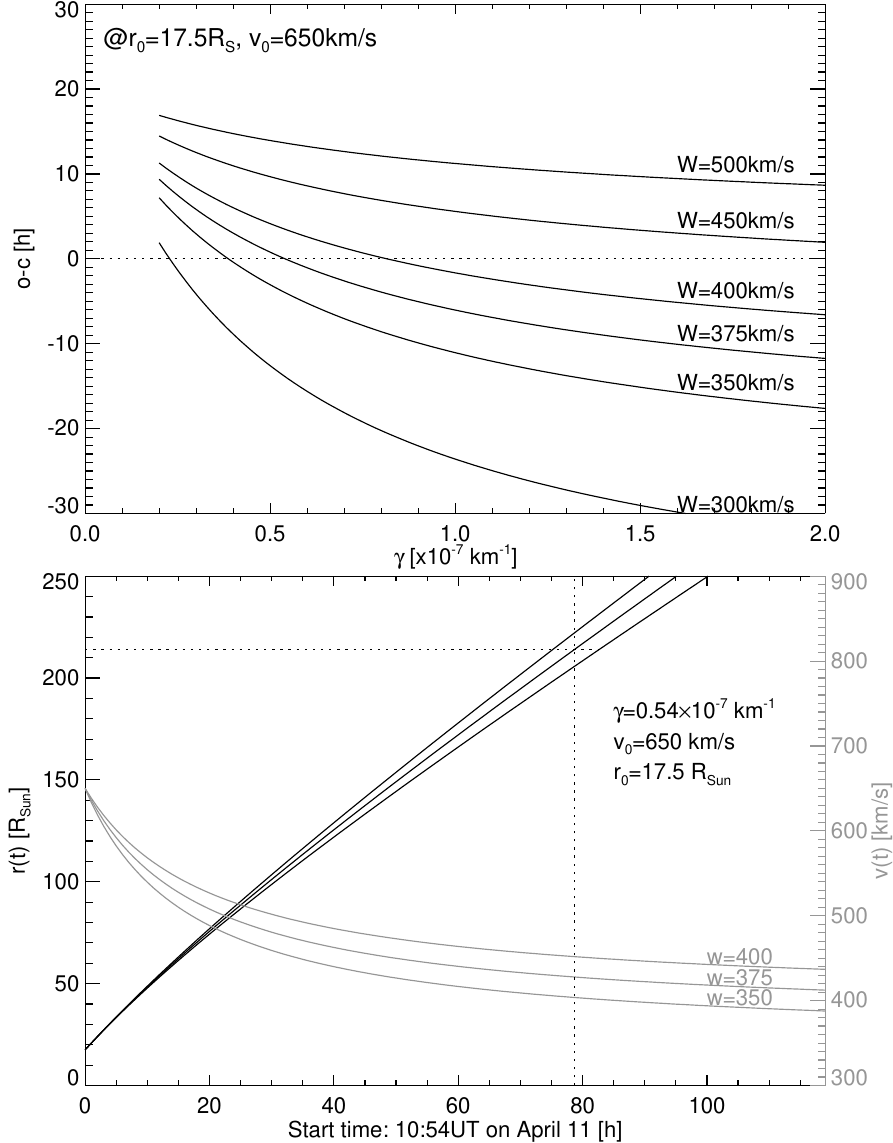}
\caption{{\bf top:} Analysis of the difference between the observed and calculated  travel times (o-c). Travel times are computed at different $w$ (indicated for each curve) as a function of $\gamma$. CME take off position $r_0=17.5R_{Sun}$ and speed $v_0=650km/s$ are considered from GCS kinematics. Intersecting point of the each curve on the horizontal (dotted) line gives $\gamma$ value for the CME arrival at L1. {\bf bottom:} CME trajectories ($r(t)$) as a function of time for ambient solar wind velocity w=350, 375 , 400 km/s. Velocity ($v(t)$) is also plotted with y-axis scale on the right. Dotted lines points the in-situ MC arrival time (April 14, 2013 at 17:35UT) on the trajectories. Note that the arrival times are differed by 5h for the range $350<w<400$. }\label{Fig15}
\end{figure}

\subsection{CME trajectory reconstructed with Drag-Based-Model}
Despite using the SSSE method with $\lambda$ equal to $90^o$, a rise in the CME speed around 14:00 UT on 2013 April 12 at a distance of $160R_\odot$ is noticed. This late increase is unphysical, for the reasons mentioned earlier and also due to the limitations of the SSSE method described in \citet{mishra2014a}. Moreover, the average in situ measured speed of the CME is around 450 km/s whereas each method (even single spacecraft based) applied in our study overestimates the speed of the tracked feature of the CME and is why these methods estimate the arrival well before the in-situ arrival. Therefore, beyond a certain radial distance from the Sun over which the CME is expected to maintain its deceleration, we have applied the drag based model (DBM; \citealt{vrsnak2013}) to estimate the arrival time and speed of the CME at L1. The model is based on the hypothesis that beyond a certain distance the CME dynamics becomes governed solely by the interaction of the CME and the ambient solar wind via aerodynamic drag \citep{cargill2004, vrsnak2010}. This assumption relies on the fact that in the interplanetary space fast CMEs decelerate, and slow ones accelerate, showing a tendency to have their velocity trend towards that of the ambient solar wind. CMEs fulfill this assumption generally at heliocentric distances around or beyond $20R_\odot$ \citep{vrsnak2004}. In the DBM model, the solution (position $r(t)$ and velocity $v(t)$) to the equation of motion describing the dynamics of the CMEs under the assumption of constant solar wind speed ($w$) and drag parameter ($\gamma$) is given by \citep{cargill2004, vrsnak2013}.

\begin{eqnarray}
 r(t)=\frac{1}{\gamma }\ln [1+\gamma \left( {{v}_{0}}-w \right)t]+wt+{{r}_{0}} \\ 
 v(t)=\frac{{{v}_{0}}-w}{1+\gamma \left( {{v}_{0}}-w \right)t}+w
\end{eqnarray}
where $r_0$, $v_0$ are initial heliocentric distance, speed of the CME.

Based on earlier works \citep{lepping1990, howard2012}, we assume that the MFR corresponds to the magnetic cloud in in situ observations and therefore, it is logical to compare the actual arrival of the magnetic cloud (here April 14, 17:35UT) with that deduced from the propagation of the MFR in COR images. As $w$ and $\gamma$ are unknown under which conditions the CME is propagating (see Equation~3), using the GCS kinematics as input to DBM, we performed a parametric analysis to obtain their approximate constrained values. 

From the GCS fitting shown in Table~\ref{tab2}, we constructed radial distance profile by including more points. We fit that profile to a second order polynomial and then derived the velocity profile by invoking a smooth cubic spline interpolation procedure. This results in an expected deceleration of the CME in its outward propagation from $4R_{\odot}$ onward. From the heliocentric distance of $17.5R_{\odot}$ at the time instance 10:54UT, we find that the CME is moving at a speed of $650km/s$. We assume from this distance onwards that the further CME propagation has no influence of Lorentz force and has not deflected because of interaction with another CME.  

Supplying $r_0=r(t=10:54UT\,on\,11)=17.5R_{\odot}$, $v_0=v(t=10:54UT\,on\,11)=650km/s$ (obtained from GCS fit) to Equation~3, we calculated the time difference between calculated (c) and observed (o: in-situ) arrival time of CME at $r(T=17:35UT\,on\,14)=214R_{\odot}$, by varying $\gamma$ at a range of w values. This time difference (o-c) as a function of $\gamma$ is plotted in Figure~\ref{Fig15}(top). Note that the curves for fast ($w>400km/s$) and slow ($w<350km/s$) solar wind speed intersect the zero time difference line at $\gamma$ values which are unsuitable for this medium mass CME \citep{vrsnak2013} and therefore the values for $\gamma$ should lie for the curves $350<w<400km/s$. Statistical studies have shown that while implementing DBM, the choice of $\gamma$ has lesser effect than ambient solar wind speed in determining the arrival time and transit speed of CMEs at 1AU \citep{vrsnak2007, vrsnak2013}. So, assuming a constant value for $\gamma$ ($=0.54\times10^{-7}km^{-1}$ which is the intercept for the curve at $w=375km/s$) within this range of solar wind, we plotted in Figure~\ref{Fig15}(bottom) the CME trajectory (Equations 3 and 4) for its initial take off position and speed obtained from GCS fit. As can be noted from this plot, the arrival times are differed only by 5hrs about the mean value curve of $w=375km/s$. Also, the transit velocities vary from mean value 428km/s by only 24km/s. This parametric analysis constructs the CME kinematics reasonably well, once we admit significant uncertainties in the CME takeoff position and speed.

\begin{figure}[h!]
\centering
\includegraphics[width=.49\textwidth]{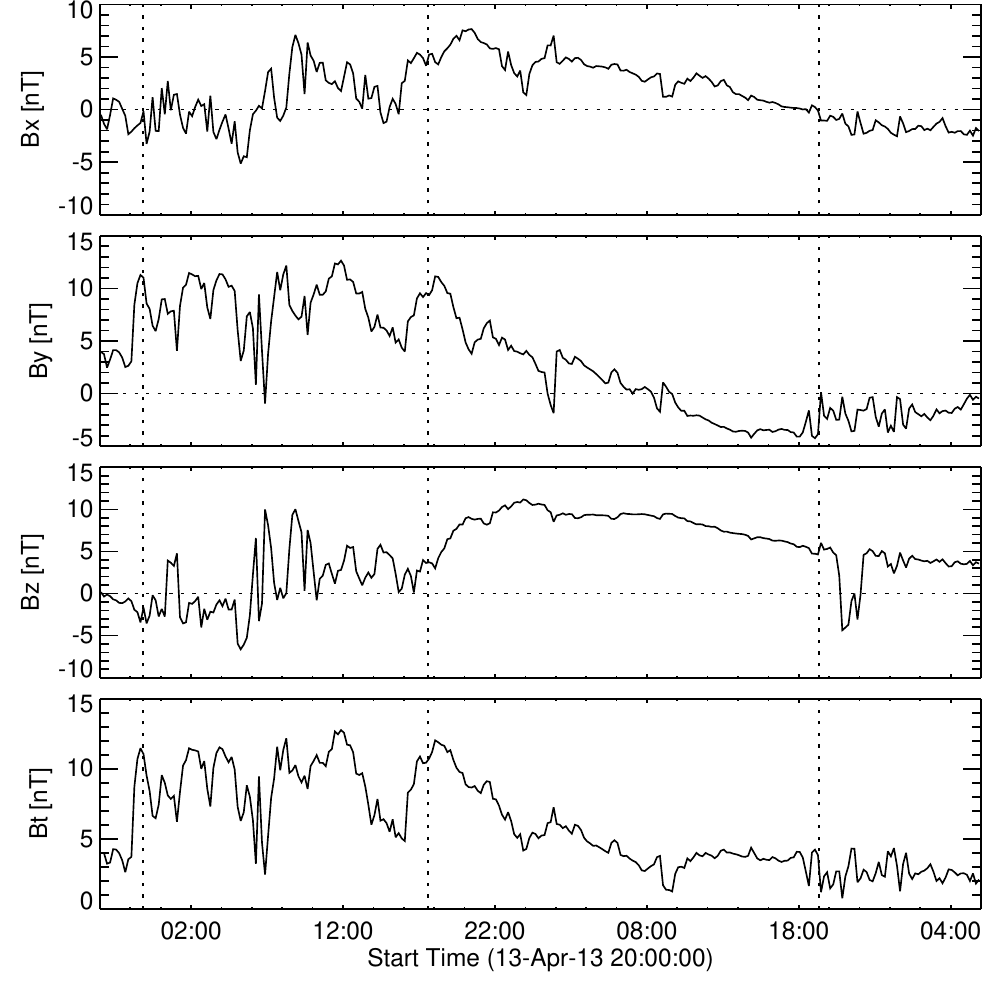}
\caption{One-minute averages of the in situ magnetic field of this CME associated MC near Lagrange point L1.  Field measurements are in GSE coordinate system. Note that $B_y$ component of the MC changes sign while passing the spacecraft. Vertical lines, from left to right, mark the arrival of  the CME shock, leading and trailing edge of the MC, respectively.}\label{Fig16}
\end{figure}

\begin{figure}[h!]
\centering
\includegraphics[width=.49\textwidth]{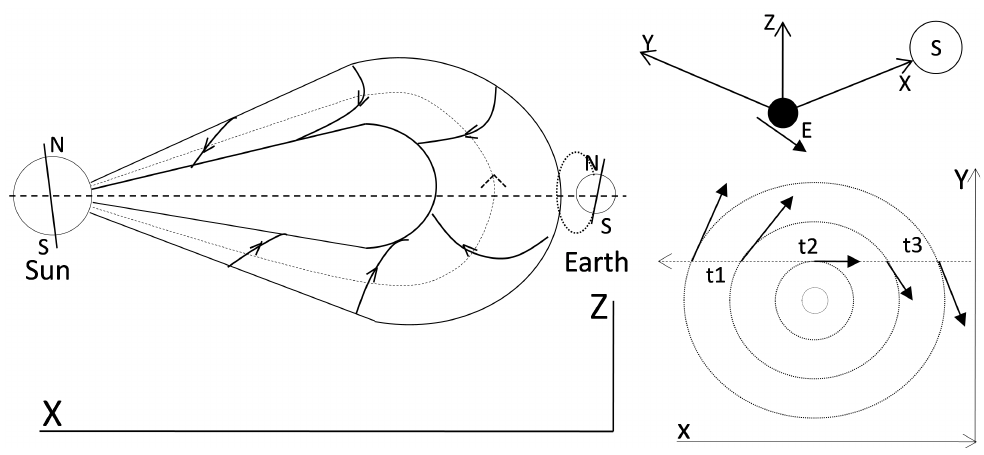}
\caption{Schematic of the MFR orientation in par with in-situ magnetic field observations. Rotation of observed $B_y$ component is demonstrated by azimuthal field and its handedness in the MC, which is shown by cross-section of MFR in XY-plane of the GSE coordinate system. Note that axial field contributes to positive $B_z$ component (northward directed) consistent with in-situ observations.}\label{Fig17}
\end{figure}
\subsection{Orientation of the magnetic field in MC of MFR} 
The source AR twist signatures are also explored in in-situ observations of MC. When a CME reaches Earth, the magnetic field strength associated with the MC is stronger than the ambient field. Depending on the orientation of the MFR,  the components of MC magnetic field vary in time. We plotted, in Figure~\ref{Fig16}, the observations of magnetic field components by WIND instrument situated near Lagrange point of Earth. These are one-minute averages and are in Geocentric Solar Ecliptic (GSE) coordinate system, where the X-axis points from Earth toward the Sun, Y is in the ecliptic plane but negative in the direction of planetary motion, and Z is parallel to the ecliptic North pole. Before the MC arrival, the field components exhibit rapid fluctuations due to shock and sheath regions. When MC passes, the magnetic field components show systematic variation indicating strong magnetic field associated to MFR. $B_x$, $B_z$ components remain positive, while $B_y$ component changes sign from positive to negative around the mid time (around 07:00UT on April 15) of the MC passage. Based on this information and the nature of the twist in the source AR, we interpreted the possible orientation of the MFR upto L1 \citep{burlaga1988, yurchyshyn2001a}.

In Figure~\ref{Fig17}, we schematically sketched the Sun-Earth connections by MFR in XZ plane of the GSE system. In this figure, axis of the MFR is approximately in meridian plane whereas the MFR is having around $-45^o$ tilt angle in the source AR. We point that there is now substantial observational evidence for the rotation of MFRs during their dynamic evolution and propagation \citep{green2007}. Our schematic, however, is motivated by the observations of MFR rotation by about $45^o$ depending on its magnetic chirality \citep{lynch2009}. In our case, although the tilt can be upto $-55^o$ in COR FOV, it can even smoothly vary upto $-90^o$ tending toward the meridian plane. In a majority of cases (~64\%), CMEs in COR FOV do have their orientation angles differed within $\pm45^o$ with their interplanetary counterparts \citep{yurchyshyn2007}.   

The axial field is northward directed (Ecliptic North) contributing to the $B_z$ component, which is positive. In this predicted orientation, the MFR cuts the XY plane in a circle having azimuthal field information and magnetic helicity (handedness) signatures of the MC. When MC passes, different regions of this cross-section encounter the spacecraft at different time phases (say t1, t2, t3 sequentially, here t2 is 07:00UT on April 15). The projection of the azimuthal field on y-axis at those time instances explains the sign and magnitude of the observed $B_y$ component. We can suitably fix the spacecraft position on the y-axis depending on the observed sign of $B_x$ component, which is positive here. Obviously, the handedness of this azimuthal field should be negative (left) in order to match the observed $B_y$-component variation from positive to negative, which is consistent with the source AR signatures of inverse S-sigmoidal structure and the negative value of $\alpha_{av}$. This event is similar to the one observed on 2000, February 21 \citep{yurchyshyn2001a} with right hand twist, where the $B_y$ component of MC changes sign from negative to positive value. It is possible to explore the twist quantitatively by reconstructing the magnetic field components in the MC cross section \citep{huq2014}, however those details are beyond the focus of this paper and requires a separate study.

\section{Summary}
\label{sec8}
CME eruptions from source active regions having soft x-ray sigmoids or H$\alpha$ filaments are modelled as manifestations of MFRs to describe many observational features. In the context of Sun-Earth connections of an eruption event on April 11, 2013, we investigate the formation/develoment scenario of a sigmoid/MFR, initiation of rise motion and its propagation toward Earth. From EUV observations of AIA, the pre-eruptive source AR consisted of a filament channel stacked over by a faintly visible inverse-S sigmoidal structure since past three days. In view of the description of this flux system as an MFR of field lines wound about some common axis, the morphological study of its evolution during 48 hour period implies a scenario of augmenting MFR. This augmentation evidently found to occur by reconnection of inclined loops lying side by the filament. Under the slow flux motions, both the transformation of distance overlying loops toward sheared arcade (stage 1), and their reconnection (stage 2) at the middle of the sigmoid are the suggested scenarios involved in the development of this sigmoidal MFR. 

The HMI magnetic fields measurements in the source AR support the EUV observations, showing monotonic decreasing net flux for the last two days. The net flux from the rope leg and the entire AR, according to MFR insertion models \citep{bobra2008, savcheva2009}, implies a marginal stability of the system of axial flux confined by the overlying polaidal flux. The average twist of the flux system in the AR also suggests a rapid build up of the MFR structure past 7 hours to the onset of the eruption. It also suggests the availability of critical twist (VZ14) in order to initiate rise motion of the MFR. Therefore, this study of EUV and magnetic fields suggests that the developed MFR system is initiated to upward motion by kink-instability to reach to a height (31Mm) from where steep gradients of horizontal field (torus-instability criteria, \citep{torok2005} drives its further outward motion in the outer corona. 

While expanding in the extended corona (after 07:10UT, 2-30$R_{\odot}$), the CME morphology captured from COR1, COR2, and LASCO views, visually fits with a parametrized MFR orientation, which aligns with the orientation of magnetic neutral line in the source active region. Tilt angle is a crucial parameter in determining orientation of MFR. MFRs by virtue of inherent twist, are shown to rotate upto $50^o$ while propagating (2-4$R_{\odot}$) in the low corona \citep{lynch2009}. This backsided halo CME did not allow us to detect the exact extent of possible MFR rotation. However, a tilt angle of $-45^o$ projects the MFR on to the plane of sky to sufficiently map the white light features of this halo CME event. An unchanged value of aspect ratio (ratio of minor to major radius of MFR=0.45) at successive stages of this CME MFR indicates its self-similar nature of expansion. Tracking methods found that the CME is propagating about $10^o$ to the East of the Sun-Earth line. This is consistent with the solar source location of the CME and the estimated direction in COR FOV from the GCS model. 

Although source AR magnetic configuration defines the MFR orientation, due to bulk in size combined with projection effects, the exact orientation of MFR in the interplanetary space is difficult to predict within $45^o$ \citep{yurchyshyn2007}. The predicted MFR orientation (MFR axis is in a vertical plane to ecliptic, see Figure~\ref{Fig17}) identifies the source AR twist signatures, which is left handed, in the in situ magnetic field observations of MC. This predicted orientation based on in situ observations, if correct, alternatively suggests a possible rotation (clockwise as seen in line-of-sight direction, to align with meridian plane) of MFR apex upto $45^o$ while traveling from Sun to the near-Earth environment. It is worth pointing that the $B_z$ component of this MC is northward, consequently no geomagnetic activity was associated with this CME.

Our analysis underlines the overestimation of the speed from all the stereoscopic methods applied on 2013 April 11 CME. Therefore, for the selected CME, when the STEREO is behind the Sun, tracking of CME shock, leading edge, and flux ropes and predicting their arrival time at 1 AU is challenging. Further, for the position of STEREO between February 2011 and 2019 June, any Earth-directed CME will be moving away from the STEREO space craft and hence, the estimation of kinematics and arrival time of CME during this time is expected to result in large errors. 

Stereoscopic reconstruction methods lead to worse results than single spacecraft reconstruction methods. We note that fixed-phi fitting method seems to perform best among all the reconstruction methods used in our study. As the physical deceleration of the CME is translated as geometrical deceleration in fitting methods, the SSEF and the HMF method seems to work poorer than the FPF method, for this quickly decelerating CME. Based on our analysis limited to a single event, we suggest that the use of FPF method should be done for practical purpose of arrival time prediction of such a CME. However, in the absence of fitting methods, we must use the reconstruction methods (i.e. TAS and HM) which approximates the CME as a very wide structure (i.e. higher value of $\lambda$) for estimating the kinematics and arrival time of CME moving at higher angle from Sun-spacecraft line. 

As almost all applied reconstruction methods failed to estimate accurate arrival time, alternatively, our study suggests that the estimated kinematics by the implementation of GCS model on COR2 observations can preferably be used as inputs in the DBM \citep{vrsnak2010} to reconstruct the CME trajectory and thereby estimate the arrival time. By a combination of parametric study and the observations, the unknown parameters like solar wind speed and drag parameter can be constrained suitably to match the in-situ arrival. 

\acknowledgements We are grateful to the referee for a detailed list of constructive comments and suggestions. SDO is a mission of NASA's Living With a Star Program, STEREO is a third mission in NASA's Solar Terrestrial Probes program, and SOHO is a mission of international cooperation between ESA and NASA. The authors sincerely thank Prof. A. Vourlidas, Prof. J. Zhang, Prof. N. Srivastava for their comments and suggestions on an earlier version of this manuscript which undoubtedly brought to its present stage. P.V. is supported by an INSPIRE grant of AORC scheme under the Department of Science and Technology.  W.M is funded by Chinese Academy of Sciences President's International Fellowship Initiative (PIFI) Grant No. 2015PE015, NSFC Grants No. 41131065 and 41574165.



\bibliographystyle{apj}


\end{document}